\documentclass{article}

\usepackage{PRIMEarxiv}

\usepackage[utf8]{inputenc} 
\usepackage[T1]{fontenc}    
\usepackage{hyperref}       
\usepackage{url}            
\usepackage{booktabs}       
\usepackage{amsfonts}       
\usepackage{nicefrac}       
\usepackage{microtype}      
\usepackage{lipsum}
\usepackage{fancyhdr}       
\usepackage{graphicx}       
\graphicspath{{media/}}     
\usepackage{float}
\usepackage{placeins}
\usepackage{multirow}
\usepackage{array}
\newcolumntype{P}[1]{>{\centering\arraybackslash}p{#1}}
\newcolumntype{M}[1]{>{\centering\arraybackslash}m{#1}}
\usepackage{graphicx}
\usepackage{amsmath}
\usepackage{amssymb}
\usepackage{float}
\usepackage{tcolorbox}
\usepackage{xcolor}

\title{Physics-Informed Feature Fusion and Structural Metadata Integration for Transferable Post-Earthquake Damage Classification: Experimental Evaluation and Community-Recovery Implications}

\author{
  Huangbin Liang\\
  Department of Civil Environmental \\ and Geomatic Engineering \\
  ETH Zürich \\
  Zurich, Switzerland\\
   \And
  Hanqing Zhang \\
  Department of Civil Environmental \\ and Geomatic Engineering \\
  ETH Zürich \\
  Zurich, Switzerland\\
  Corresponding email: hanqingtju@gmail.com\\
  \AND
  Jiazeng Shan\\
  Department of Civil Engineering \\
  Tongji University\\
  Shanghai, China\\
  \And
  Eleni Chatzi \\
  Department of Civil Environmental \\ and Geomatic Engineering \\
  ETH Zürich \\
  Zurich, Switzerland\\
}

\begin{document}
\maketitle

\begin{abstract}
Earthquake-induced structural damage assessment remains a fundamental challenge for population-based Structural Health Monitoring (PBSHM), where generalizable damage representations across structurally heterogeneous buildings are critically needed. Existing vibration-based approaches commonly rely on structure-specific damage-sensitive features (DSFs) and often adopt uniform code-prescribed drift thresholds for damage labeling, limiting their generalization capability across diverse structural systems. This study proposes a physics-informed feature fusion and structural metadata integration framework for robust post-earthquake structural damage classification and resilience-oriented assessment. A large-scale nonlinear simulation dataset is generated using building populations with varying geometrical and material configurations and subjected to multiple earthquake scenarios. To ensure physically consistent damage labeling, structure-specific damage states are adaptively defined through nonlinear pushover analysis and capacity-based threshold interpretation rather than fixed drift limits. Based on sparse ground and roof acceleration measurements, a suite of physics-informed DSFs is formulated and systematically compared against Catch22 statistical descriptors and convolutional kernel-based MiniRocket representations under group-wise cross-structure validation. Multiple machine-learning classifiers are employed to evaluate feature robustness and reduce model-specific bias. Results show that physics-informed DSFs outperform generic time-series representations in transferable damage classification tasks under the considered sparse-sensing setting. Progressive enrichment with structural metadata further improves classification robustness, with geometrical and material information providing complementary structural context for interpreting response-based DSFs. Modal information contributes only marginally in the simulated dataset, but substantially improves performance in out-of-distribution shaking-table validation, highlighting its role in bridging simulation-to-experiment discrepancies. Finally, inferred damage states are integrated into a community-level recovery model, showing that monitoring-informed damage tags can reduce inspection delay, improve repair prioritization, and reduce resilience loss across different seismic intensity levels. The findings highlight how transferable feature design and information fusion can convert monitoring data into actionable information for post-earthquake damage assessment and resilience-oriented decision-making.

\end{abstract}


\keywords{Physics-informed damage-sensitive features \and cross-structure damage classification \and structural metadata integration \and post-earthquake damage assessment \and population-based structural health monitoring.}

\section{Introduction}
Earthquakes continue to pose significant threats to the safety, functionality, and resilience of civil structures and infrastructure systems \cite{mieler2018review,sun2019resilience}. Although modern seismic design philosophies aim to prevent catastrophic collapse, performance-based seismic design inherently tolerates varying levels of structural damage and functionality loss under strong ground motions \cite{choudhury2024performance}. The rapid quantification of earthquake-induced damage, often represented through categorical damage states (DSs), is therefore essential for supporting emergency response, occupancy decisions, and recovery planning following seismic events \cite{anwar2023systems,liang2025resilience}. However, timely and reliable assessment becomes increasingly challenging when large numbers of structures must be inspected following major seismic events. Conventional post-earthquake assessment procedures predominantly rely on visual inspections and engineering judgment. While widely adopted in practical engineering, such approaches are labor-intensive, time-consuming, and difficult to scale to large building inventories \cite{khakurel2023post, galloway2014lessons}. Moreover, visual inspections can be unsafe due to falling hazards and suffer from subjectivity among inspectors, particularly for structures exhibiting moderate or distributed damage patterns. These limitations have motivated the development of automated and monitoring-based approaches capable of supporting rapid post-earthquake condition assessment.

Structural Health Monitoring (SHM) provides a promising framework for inferring structural condition and earthquake-induced damage from measured structural responses \cite{farrar2007introduction}. Since direct measurements of internal damage mechanisms are generally impractical for large-scale building inventories \cite{kaya2015real}, vibration-based SHM approaches have attracted significant attention, and they aim to identify damage indirectly through changes in global structural dynamics associated with stiffness degradation and nonlinear structural behavior \cite{zar2024towards}. Recent advances in sensing technologies, low-cost accelerometers, wireless acquisition systems, and Internet-of-Things infrastructures have further increased the feasibility of permanent vibration monitoring deployments in civil structures \cite{hannan2022review}. Based on the relationship between structural degradation and dynamic response variation, numerous vibration-based damage-sensitive features (DSFs) have been proposed for earthquake-induced damage identification, including modal-property-based indicators \cite{pandey1991damage,shokrani2018use}, transmissibility-based features \cite{liu2023data,luo2021weighted}, autoregressive coefficients \cite{yao2012autoregressive,chegeni2022new}, spectral descriptors \cite{yang2017fourier}, wavelet-energy metrics \cite{balafas2015development}, and time-domain statistical features \cite{hannan2022review}. For example, Shokrani et al. \cite{shokrani2018use} proposed a damage localization method leveraging mode shape curvatures combined with PCA to eliminate the effects of environmental/operational variability using response-only data. Wen et al. \cite{wen2025non} developed a frequency-domain method based on local damping parameter screening to automatically identify earthquake-induced micro-damage in substation equipment. Liu et al. \cite{liu2023data} applied power spectral density transmissibility functions of strain responses combined with inverse Fourier transformation to identify structural damage in beam-like structures under unknown seismic excitations. Chegeni et al. \cite{chegeni2022new} adopted autoregressive model coefficients and residuals extracted from time series analysis to localize and quantify structural damage. Among them, modal features extracted through operational modal analysis (OMA), such as changes in natural frequencies, damping ratios, and mode shapes, have been extensively investigated and widely applied due to their physical interpretability and direct relationship to structural stiffness degradation \cite{shokrani2018use,pooya2021novel,vidal2014changes,astroza2022statistical,wen2025non}. However, most existing DSFs are sensitive to specific structural configurations, dynamic characteristics, damage mechanisms, excitation types, and noise conditions of the measurement from which they are derived. Consequently, a single DSF that performs well for one structure may exhibit significantly degraded performance when applied to structurally different buildings \cite{reuland2021damage,martakis2023fusing,zhang2024post}. This limitation becomes particularly critical in the context of Population-Based Structural Health Monitoring (PBSHM) \cite{bull2021foundations,astorga2025exploring}, where the objective is not merely achieving accurate damage classification for an individual structure, but rather developing transferable damage representations capable of robustly generalizing across heterogeneous structural populations with varying geometries, material properties, and dynamic characteristics.

To overcome the limitations of individual DSFs, machine learning (ML) techniques have increasingly been introduced to fuse multiple features into automated damage classifiers \cite{farrar2012structural}. Supervised learning algorithms, such as Support Vector Machines, Random Forests, Gradient-Boosted Decision Trees, and Neural Networks, have demonstrated promising capabilities for multi-class structural damage classification using feature combinations \cite{harirchian2021synthesized,martakis2023fusing,lu2021deep, bhatta2024machine}. For instance, Bhatta and Dang \cite{bhatta2024machine} applied multiple machine learning classifiers using combined structural properties and ground motion characteristics to classify post-earthquake building damage at a regional scale. Lu et al. \cite{lu2021deep} proposed a CNN-based rapid seismic damage assessment method using time-frequency distribution graphs of ground motions as input features to predict building damage states in near real-time. Nevertheless, many existing studies primarily rely on structural properties or seismic intensity measures \cite{lu2021deep,bhatta2024machine,stojadinovic2022rapid,zhu2024post} while only limited attention has been devoted to the integration of physics-informed SHM features derived from SHM vibration responses \cite{martakis2023fusing}. More recently, generic time-series representation methods, such as Catch22 descriptors \cite{lubba2019catch22} and ROCKET-based transformations using random convolutional kernels \cite{dempster2019rocket,dempster2021minirocket}, have emerged as powerful alternatives capable of automatically extracting discriminative patterns directly from raw signals without relying on manually designed structural features. These methods have achieved remarkable performance in general time-series classification tasks \cite{tan2022multirocket} and are increasingly attracting attention within SHM applications \cite{avci2021review}. Nevertheless, systematic comparisons between individual and fused DSFs are limited, and it remains unclear whether generic time-series representations can achieve better transferability than physics-informed DSFs for PBSHM across structurally heterogeneous systems.

Several recent studies have explored transfer-learning and domain-adaptation strategies to improve the cross-domain generalization capability of damage classification models \cite{Zhang2025Transfer,LIU2024Crossdomain,LI2024118928,poole2023statistic}. However, many existing frameworks still rely on the availability of at least limited target-domain monitoring data for model adaptation or feature alignment \cite{Zhang2025Transfer,poole2023statistic}. Moreover, the effectiveness of many existing approaches has primarily been investigated using numerical simulations, whereas experimentally validated studies involving real structural damage remain relatively limited. In practical earthquake engineering applications, such assumptions may be difficult to satisfy because reliable labeled earthquake-induced structural damage data remain extremely scarce. Consequently, identifying intrinsically transferable damage representations may provide a more fundamental pathway toward robust population-based damage classification. In this context, another promising direction involves enriching measurement-based DSFs with structural metadata, such as geometrical properties, material parameters, and modal characteristics \cite{martakis2023fusing}. Such information may provide complementary descriptions of the underlying structural system and help improve cross-structure generalization capability \cite{KUO2025113467}. Nevertheless, the relative contributions of different categories of structural metadata to transferable damage classification in PBSHM remain insufficiently understood.

Another important challenge concerns the definition of structural damage labels within heterogeneous building populations. In many previous studies, global DSs are assigned using fixed inter-story drift ratio thresholds prescribed by seismic design standards and uniformly applied across all structures \cite{mangalathu2020classifying,ding2025seismic}. However, identical drift ratios may correspond to fundamentally different structural degradation states depending on the structural configuration, ductility capacity, and nonlinear energy dissipation characteristics of the system. Consequently, fixed drift-based thresholds may introduce physically inconsistent labels that reduce the transferability of learned damage classifiers. 

Furthermore, most existing SHM and PBSHM studies remain primarily focused on isolated structural-level damage identification accuracy \cite{giordano2023quantifying,giordano2022value}, while the broader implications of transferable monitoring-based damage representations for downstream resilience and recovery analyses remain insufficiently explored \cite{straub2017value,liang2025harnessing}. In the context of resilience-oriented post-earthquake assessment and enhancement, structural DSs directly influence repair decisions, recovery trajectories, and functionality restoration of infrastructure systems \cite{makhoul2024seismic,liang2026quantifying,argyroudis2022digital}. However, relatively few studies have investigated how measurement-informed damage classification can influence post-earthquake recovery processes and overall community resilience quantitatively \cite{liang2026quantifying}. 

Motivated by these challenges, this study proposes a physics-informed feature fusion and structural metadata integration framework for robust population-based post-earthquake structural damage classification. Specifically, this study makes four contributions. First, a nonlinear simulation dataset of heterogeneous RC frame buildings is generated, and structure-specific DS labels are assigned through pushover-based capacity interpretation rather than fixed drift thresholds. Second, physics-informed DSFs are designed from sparse ground-roof acceleration measurements and systematically compared with Catch22 \cite{lubba2019catch22} and MiniRocket \cite{dempster2021minirocket} under group-wise cross-structure validation. Third, the contribution of structural metadata is examined by progressively adding geometrical, material, and modal information to the response-based feature space. Fourth, the simulation-trained classifiers are evaluated using out-of-distribution shake table test data and further integrated into a community recovery model to quantify the downstream resilience value of SHM-informed damage tagging.

The findings indicate that physics-informed DSFs provide more transferable and robust representations for cross-structure post-earthquake damage classification than generic time-series representations. Structural metadata is further shown to provide complementary information that improves classification robustness across heterogeneous structural populations. Interestingly, while modal properties provide only marginal improvement in purely simulated datasets, they significantly enhance predictive performance in experimental settings, highlighting the strong influence of data fidelity, measurement uncertainty, and simulation-to-reality discrepancies on transferable damage representations. In addition, the integration of measurement-informed damage tagging into community-level recovery assessment demonstrates the potential of SHM-enhanced damage identification to support faster recovery and improved post-earthquake resilience. Overall, this study advances the understanding of transferable vibration-based damage representations for PBSHM and provides new insights into how physics-informed feature engineering and structural metadata enrichment can support robust post-earthquake damage assessment and downstream resilience-oriented decision-making.

Accordingly, the study is organized around four linked research questions: (i) whether physically motivated response features are more transferable across heterogeneous structures than generic time-series representations; (ii) whether structural metadata improves the interpretation of response-derived damage indicators; (iii) whether the resulting representations retain predictive value when transferred from simulation to an external experimental structure; and (iv) how uncertainty in monitoring-informed damage classification propagates into an illustrative community-recovery analysis. The first three questions concern damage-representation transferability, whereas the fourth examines the downstream decision relevance of the inferred damage states rather than introducing a new recovery-optimization method.

\section{Framework}
\begin{figure}[!htbp]
    \centering
    \includegraphics[width=1\linewidth]
    {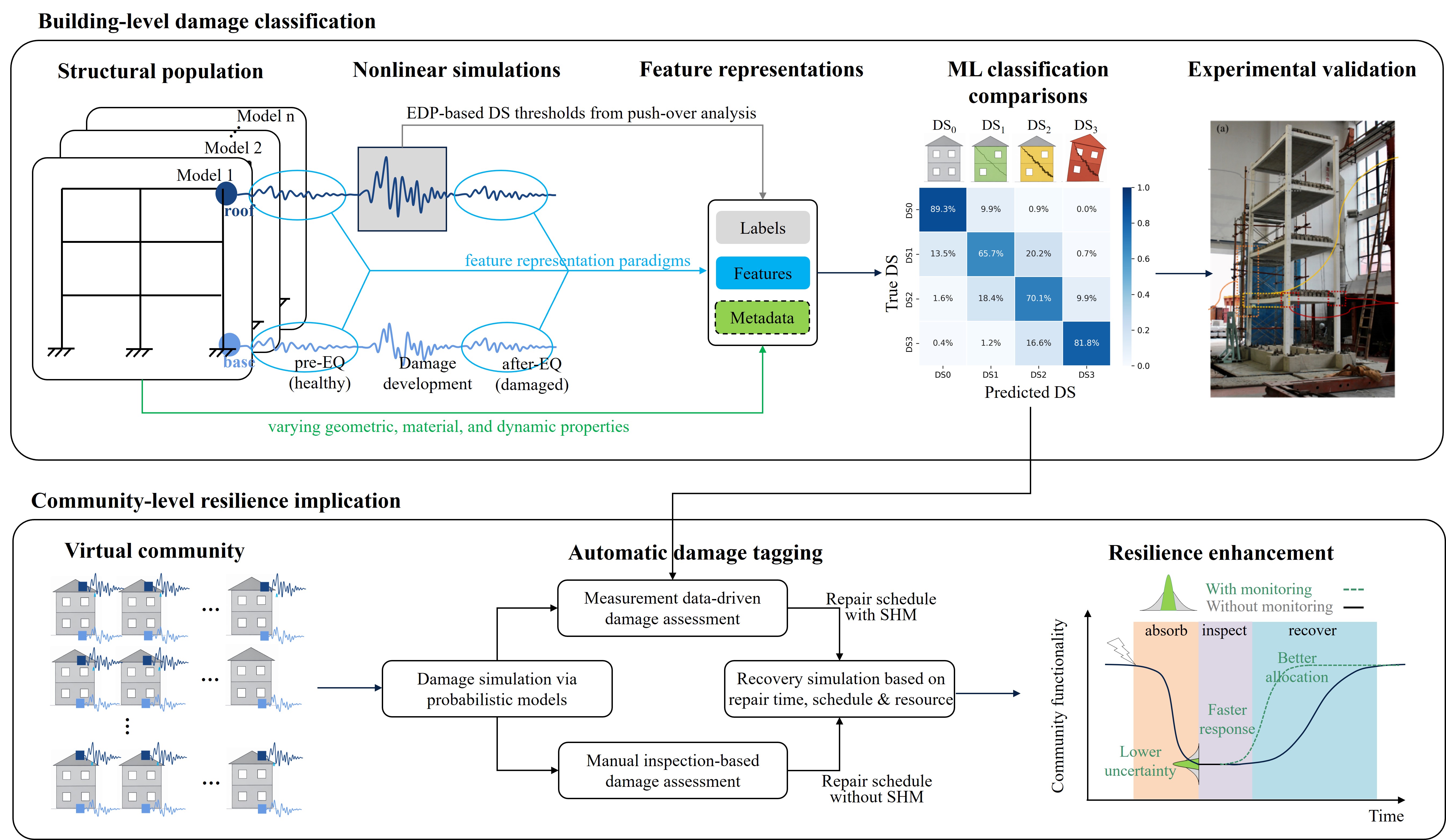}
    \caption{Overall framework for transferable population-based post-earthquake structural damage classification and resilience-oriented assessment. A parametric population of building structures is first generated with varying geometrical, material, and dynamic properties. Simulated seismic responses are used to construct different feature representation paradigms, including physics-informed DSFs, generic statistical time-series descriptors, and automatic kernel-derived representations. These representations are combined with physically consistent damage labels and structurally enriched feature sets, and evaluated through ML-based multi-class damage classification. The transferability of the learned representations and robustness of the trained ML classifier are further examined using experimental shake table data. Finally, monitoring-informed damage assessment results are integrated into a community-level recovery model to compare post-earthquake resilience under different information scenarios, including cases with and without SHM.}
    \label{framework}
\end{figure}

The proposed framework is designed to investigate transferable damage representations for population-based post-earthquake structural damage classification, as illustrated in Figure \ref{framework}. The central objective is not only to achieve accurate damage identification for individual structures, but also to examine whether the learned damage representations and trained ML classifier can generalize across structurally heterogeneous building populations and enhance community resilience. To this end, the framework integrates structural population generation, nonlinear seismic response simulation, damage-sensitive feature extraction, generic time-series representation, structural metadata enrichment, ML-based damage classification, experimental validation, and resilience-oriented recovery assessment, as detailed in the following sections.

\subsection{Structural population generation}
A parametric population of two-dimensional reinforced concrete (RC) frame structures was generated to construct the numerical dataset for population-based damage classification. Structural variability was introduced through randomized sampling of geometrical, material, loading, and dynamic parameters within predefined engineering ranges, as summarized in Table \ref{tab:1}. A total of 100 parametric structural models were generated, and finite-element models corresponding to each parameter set were automatically established via OpenSeesPy \cite{zhu2018openseespy} for subsequent nonlinear pushover and seismic response analyses. 

The generated frames include 3–7 stories and 2–4 bays. Story heights and bay widths were randomly sampled between 3.0–3.6 m and 4.5–7.0 m, respectively. Beam and column cross-sectional dimensions were also varied to generate structures with different stiffness and strength characteristics. The column section width and depth were sampled from 0.45–0.70 m, while the beam section width and depth varied between 0.25–0.40 m and 0.40–0.75 m, respectively. Material and loading properties were further randomized to represent variability commonly observed in real building inventories through random sampling of concrete compressive strength, steel yield strength, and floor loads. Concrete compressive strength ($f_c$) was sampled within 25–40 MPa, while reinforcing steel yield strength ($f_y$) varied between 300 and 400 MPa. The concrete elastic modulus ($E_c$) was computed from the sampled compressive strength according to the ACI empirical relation ($E_c = 4700\sqrt{f_c}$) (MPa), while the steel elastic modulus ($E_s$) was fixed at 200 GPa. Dead and live loads were sampled from 3–6 $kN/m^2$ and 1–4 $kN/m^2$, respectively. For the two-dimensional frame idealization, the tributary width was taken as the bay width, and the floor mass was computed from the dead load and 60\% of the live load.

Based on the sampled parameters, each frame was modeled using nonlinear beam-column elements with fiber sections in OpenSees \cite{mckenna2011opensees}. Concrete was represented using the Concrete02 material model, while reinforcing steel was modeled using Steel02 with strain limits imposed through a MinMax wrapper. Rectangular RC fiber sections were generated for beams and columns using confined core concrete, cover concrete, and longitudinal reinforcement layers. A constant longitudinal reinforcement ratio of 1\% was adopted for both beams and columns, representing a typical reinforcement level for ordinary reinforced concrete frame buildings and allowing the influence of geometric, material, and loading variability to be isolated without introducing additional uncertainty associated with reinforcement detailing. The confinement effect of the core concrete was represented through the Concrete02 material model, while the cover concrete was assumed to remain unconfined. Five Lobatto integration points were adopted for both beam and column elements. Column elements were assigned a P-Delta geometric transformation, while beam elements were modeled using a linear geometric transformation.

After model generation, gravity analysis was first performed to establish the initial stress state. Modal analysis was then conducted to extract the fundamental dynamic properties, including the first several natural periods and frequencies. A damping ratio randomly sampled between 4.5\% and 5.5\% was assigned to represent moderate variability in structural energy dissipation. Rayleigh damping was adopted, with the mass- and stiffness-proportional damping coefficients calibrated using the first two vibration modes to achieve the target damping ratio. These modal quantities were stored as part of the structural metadata and later used to evaluate the contribution of dynamic-property information to transferable damage classification.

\begin{table}[ht]
    \centering
    \caption{Randomized parameters for generating the population of two-dimensional RC frame structures}
    \label{tab:1}
    \begin{tabular}{l c c c c}
        \toprule
        Parameters & Denotation & Unit & Min & Max \\
        \midrule
        Number of stories        & $n_s$ & /       & 3   & 7   \\
        Number of bays           & $n_b$ & /       & 2   & 4   \\
        Story height             & $H_s$ & m       & 3.0 & 3.6 \\
        Bay width                & $B_w$ & m       & 4.5 & 7.0 \\
        Column section width     & $b_c$ & m       & 0.45 & 0.70 \\
        Column section depth     & $h_c$ & m       & 0.45 & 0.70 \\
        Beam section width       & $b_b$ & m       & 0.25 & 0.40 \\
        Beam section depth       & $h_b$ & m       & 0.40 & 0.75 \\
        Concrete strength        & $f_c$ & MPa     & 25  & 40  \\
        Steel yield strength  & $f_y$ & MPa     & 300 & 400 \\
        Dead load                & $q_D$ & kN/m$^2$ & 3  & 6   \\
        Live load                & $q_L$ & kN/m$^2$ & 1  & 4   \\
        Damping ratio            & $\zeta$ & \%    & 4.5 & 5.5 \\
        \hline
    \end{tabular}
\end{table}

\subsection{Nonlinear simulations}
The nonlinear analysis procedure consisted of three main stages, including pushover analysis for damage state threshold definition, white-noise excitation analysis under undamaged conditions, and nonlinear time-history analyses under earthquake excitation followed by post-earthquake white-noise response simulations.

\begin{figure}[!htbp]
    \centering
    \includegraphics[width=0.8\linewidth]
    {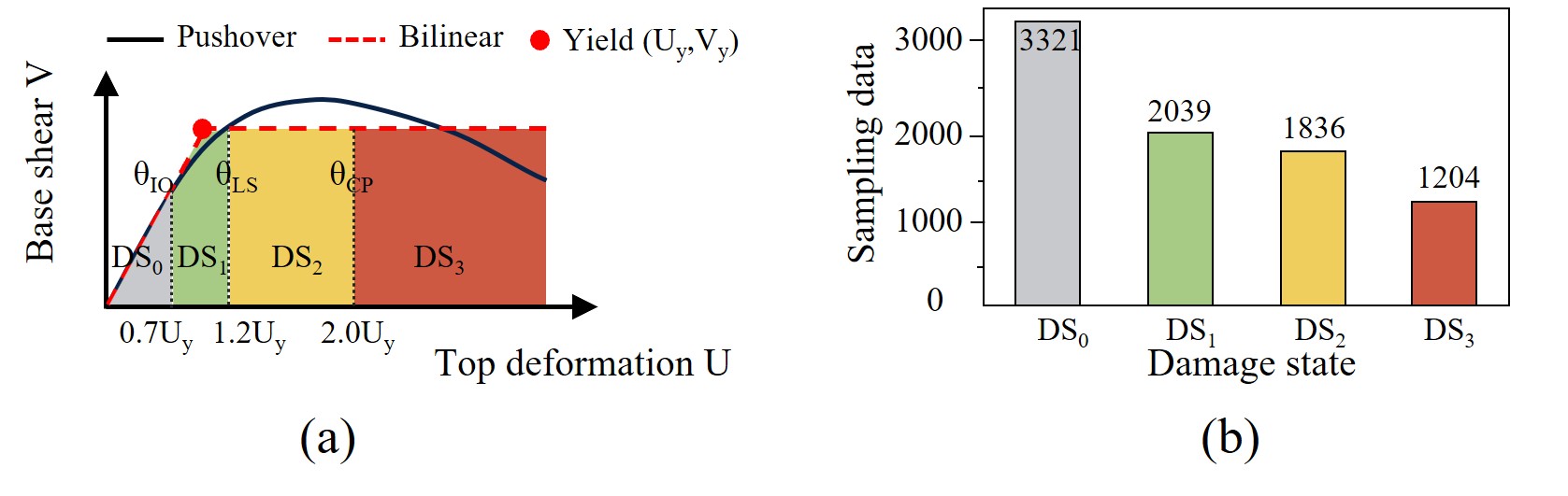}
    \caption{(a) Definition of structural DS based on pushover-derived capacity thresholds; (b) distribution of structural DS in the generated simulation dataset.}
    \label{DS_definition_distribution}
\end{figure}

For each generated structural model, a nonlinear pushover analysis was first conducted using a displacement-controlled procedure with roof displacement as the control degree of freedom. An inverted triangular lateral load pattern proportional to the story height was adopted to approximate the dominant first-mode response of regular low- to mid-rise RC frame buildings. The analysis was continued until the maximum inter-story drift ratio reached 10\%, ensuring that the complete nonlinear capacity curve, including the post-yield response, was captured. The pushover analysis yielded the global base-shear–roof-displacement capacity curve for each structure. To derive structure-specific yielding characteristics, the yielding displacement ($U_y$) was determined by enforcing equivalence between the energy dissipated by the original pushover curve and the idealized bilinear curve \cite{fajfar1999capacity}, as illustrated in Figure \ref{DS_definition_distribution}(a). Based on the identified yielding displacement, structure-specific global performance points were defined to establish damage-state thresholds. Three performance levels were considered using roof displacement limits proportional to the yielding displacement, namely (0.7$U_y$), (1.2$U_y$), and (2.0$U_y$), which are broadly associated with the Immediate Occupancy (IO), Life Safety (LS), and Collapse Prevention (CP) performance levels commonly adopted in performance-based seismic assessment frameworks \cite{ASCE41,FEMA356}. For each performance point, the corresponding maximum inter-story drift ratio ($\theta$) obtained from the pushover analysis was extracted and used as the structural damage state threshold. In this manner, the damage labels were derived from the actual nonlinear capacity of each structure rather than fixed code-prescribed drift limits, enabling consistent characterization of nonlinear structural capacity across structurally heterogeneous systems.

Following the pushover analysis, white-noise (WN) excitation analyses were first conducted for the undamaged structures to generate baseline vibration-response datasets. Band-limited white-noise acceleration with a low amplitude of 0.05g was applied at the structural base to excite the dominant dynamic characteristics of each model while maintaining an essentially linear structural response. At this excitation level, the white-noise excitation was assumed not to introduce additional structural damage or permanent stiffness degradation. The resulting acceleration responses were subsequently used as healthy reference signals for vibration-based feature extraction and comparative post-earthquake assessment.

To simulate earthquake-induced structural damage, nonlinear seismic analyses were performed using a suite of 12 earthquake ground motions selected from the European Strong Motion Database (ESD) \cite{iervolino2010rexel}. The records were selected to represent variability in earthquake characteristics while remaining compatible with the target seismic hazard level defined by the SIA 261 design spectrum \cite{SIA261}. Detailed information on the selected ground-motion records is provided in Appendix A. As illustrated in Figure \ref{EQs}, the mean response spectrum of the selected records generally envelopes the code-based spectrum within the dominant structural frequency range of the generated building population (1–8 Hz). To further increase the diversity of structural response and damage conditions, each ground motion was scaled to multiple peak ground acceleration (PGA) levels ranging from 0.2 g to 1.4 g at intervals of 0.2 g. For each simulation case, a nonlinear dynamic analysis was conducted using the Newmark integration scheme combined with Rayleigh damping calibrated from the first two vibration modes. During the analyses, displacement responses were recorded at each story level, and the maximum inter-story drift ratio $\theta$ was subsequently computed from the recorded displacement histories. The maximum $\theta$ obtained during each earthquake simulation was then compared against the previously derived structure-specific drift thresholds to assign the corresponding global DSs. After completion of each earthquake excitation, a free-vibration decay analysis was first performed to dissipate transient numerical oscillations before conducting post-earthquake vibration testing. Subsequently, an additional WN excitation analysis was carried out on the damaged structure using the same low-amplitude excitation procedure adopted for the healthy-state analysis. These post-earthquake white-noise responses were used to characterize damage-induced changes in structural dynamic behavior while minimizing the influence of transient earthquake-input characteristics. By comparing the healthy and post-earthquake vibration responses, physically meaningful damage-sensitive features could subsequently be extracted for transferable damage classification analyses.

Overall, the generated dataset consists of 8,400 nonlinear simulation cases obtained from combinations of 100 structural models, 12 earthquake ground motions, and 7 PGA scaling levels. Figure \ref{DS_definition_distribution}(b) summarizes the resulting DS distribution within the generated dataset. Although the dataset exhibits moderate class imbalance, all four DSs remain sufficiently represented for subsequent multi-class classification analyses.

\begin{figure}
    \centering
    \includegraphics[width=0.9\linewidth]
    {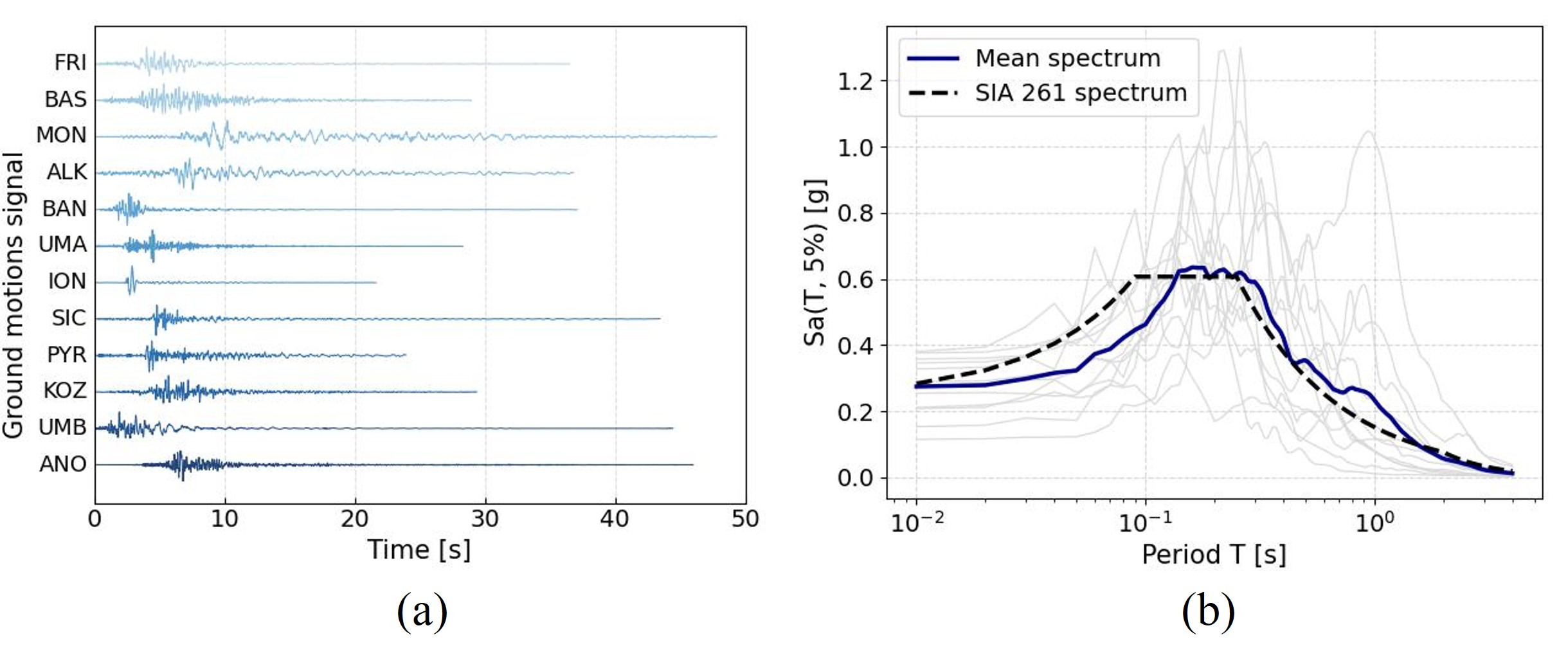}
    \caption{Selected earthquake ground motions and corresponding response spectra. (a) acceleration time histories of the 12 ground motions selected from the European Strong Motion Database (ESD) \cite{iervolino2010rexel}; (b) comparison between the individual and mean response spectra of the selected records and the target SIA 261 design spectrum \cite{SIA261}.}
    \label{EQs}
\end{figure}

\subsection{Feature representations}
To define a generic feature space independent of the individual structural configuration, all feature representations were extracted using only the ground and roof acceleration responses from the healthy-state and post-earthquake white-noise excitation analyses, as illustrated in Figure \ref{framework}. This configuration avoids dependence on the number of stories, sensor density, or structural topology of individual models, while remaining consistent with practical vibration-monitoring scenarios using sparse measurements. Based on these response signals, three representation paradigms were considered in this study, namely physics-informed DSFs, generic statistical time-series descriptors based on Catch22 \cite{lubba2019catch22}, and convolutional kernel-derived representations via MiniRocket \cite{dempster2021minirocket}. It should be noted that the three representation paradigms do not rely on identical sources of information. The proposed DSFs explicitly utilize roof–ground response relationships and structure-specific modal frequency bands derived from the healthy-state analysis, whereas Catch22 and MiniRocket operate according to their original formulations. The objective of this comparison is therefore to evaluate whether embedding structural dynamics knowledge into feature construction improves transferable damage characterization.

\subsubsection{Physics-informed DSFs}
Physics-informed DSFs were manually designed based on structural dynamics principles and physically interpretable mechanisms associated with earthquake-induced structural degradation. These features were formulated to explicitly capture changes in structural stiffness, modal characteristics, response amplification, and temporal response evolution caused by structural damage. The extracted DSFs can be broadly categorized into frequency-domain features and time-domain features.

Frequency-domain DSFs were extracted from transmissibility functions computed between the roof and ground acceleration responses. In this study, transmissibility was estimated using Welch-based spectral densities in Eq. (\ref{eq1}).
\begin{equation}
    T(f)=\frac{P_{gr}(f)}{P_{gg}(f)}, \quad f \in [f_a,f_b]
    \label{eq1}
\end{equation}
where $P_{gg}(f)$ denotes the auto-spectral density of the ground acceleration; $P_{gr}(f)$ denotes the cross-spectral density between the ground and roof acceleration responses; $[f_a,f_b]$ represents the selected frequency band of interest associated with the dominant structural modes. Based on the healthy-state and post-earthquake transmissibility spectra, i.e., $T_h(f)$ and $T_d(f)$, several frequency-domain DSFs were subsequently extracted. 
The peak-frequency-shift feature was first introduced to quantify damage-induced modal-frequency variations:
\begin{equation}
    F_{\mathrm{peak}} = 
    \frac{
        \underset{f \in [f_a, f_b]}{\arg\max}\, T_{h}(f) - 
        \underset{f \in [f_a, f_b]}{\arg\max}\, T_{d}(f)
    }{
        \underset{f \in [f_a, f_b]}{\arg\max}\, T_{h}(f)
    }
    \label{eq2}
\end{equation}
in which $\arg\max$ operator returns the frequency location corresponding to the maximum transmissibility amplitude within the selected frequency range $[f_a,f_b]$. Consequently, $F_{peak}$ represents the normalized shift of the dominant transmissibility peak caused by structural damage.

While $T_{peak}$ only captures the dominant transmissibility peak, the centroid-frequency feature additionally incorporates information from the entire modal frequency band:
\begin{equation}
    f_c =
    \frac{
    \sum_{f \in [f_a,f_b]} f\cdot T(f)
    }{
    \sum_{f \in [f_a,f_b]}T(f)
    }
    \label{eq3}
\end{equation}

The normalized centroid-frequency shift was then defined as:
\begin{equation}
    F_{\mathrm{centroid}} = 
    \frac{
    f_{c,h}-f_{c,d}
    }{
    f_{c,h}
    }
    \label{eq4}
\end{equation}
where $f_{c,h}$ and $f_{c,d}$ denote the transmissibility spectral centroids of the healthy and damaged states, respectively.

To further quantify changes in the overall transmissibility shape, the transmissibility-based modal assurance criterion (MAC) feature was computed as:
\begin{equation}
    F_{\mathrm{MAC}} = 1 -
    \frac{
    \left|
    \sum_{f \in [f_a,f_b]}
    T_h(f)\overline{T_d(f)}
    \right|^2
    }{\left(
    \sum_{f \in [f_a,f_b]}
    |T_h(f)|^2
    \right)
    \left(
    \sum_{f \in [f_a,f_b]}
    |T_d(f)|^2
    \right)
    }
    \label{eq5}
\end{equation}
where $\overline{T_d(f)}$ denotes the complex conjugate of the damaged-state transmissibility, and larger $DSF_{MAC}$ values indicate an increasing deviation of the damaged transmissibility spectrum from the healthy-state reference.

Finally, the transmissibility spectral-area variation was introduced to quantify damage-induced changes in modal-band energy distribution:
\begin{equation}
    F_{\mathrm{area}} = 
    \frac{
    \displaystyle \int_{f_a}^{f_b} |T_h(f)|\,df
    -
    \displaystyle \int_{f_a}^{f_b} |T_d(f)|\,df
    }{
    \displaystyle \int_{f_a}^{f_b} |T_h(f)|\,df
    }
    \label{eq6}
\end{equation}
where $\int_{f_a}^{f_b}|T(f)|\,df$ denotes the integrated transmissibility spectral area within the selected frequency band.
All frequency-domain DSFs were evaluated within two modal-centered frequency bands, namely $[0.4f_1,,1.2f_1]$ and $[0.4f_2,,1.2f_2]$, where $f_1$ and $f_2$ denote the first and second natural frequencies of the healthy structure, respectively. These bands were selected to focus the analysis around the dominant modal peaks while reducing the influence of irrelevant spectral components and high-frequency noise. Accordingly, each frequency-domain feature was computed separately for the first- and second-mode frequency regions, as listed in Table \ref{tab:2}.

Time-domain DSFs were extracted directly from the roof and ground acceleration response histories to characterize damage-induced changes in response delay, response amplitude, temporal response evolution, and effective dynamic stiffness.
The time-delay feature $\tau$ was computed from the cross-correlation between the ground acceleration $x(t)$ and roof acceleration response $y(t)$:
\begin{equation}
    T_{\mathrm{delay}} = 
    \left| \arg \max_{\Delta t} \left| R_{xy,d}(\Delta t) \right| - 
    \arg \max_{\Delta t} \left| R_{xy,h}(\Delta t) \right| \right|
    \label{eq7}
\end{equation}
where $R_{xy}(\Delta t)$ denotes the cross-correlation function, and the $\arg\max$ operator returns the delay corresponding to the maximum correlation magnitude. The subscripts $h$ and $d$ represent the healthy and damaged states, respectively.

The Root Mean Square (RMS) ratio feature was computed from the roof acceleration responses normalized by the RMS value of the corresponding ground-motion input:
\begin{equation}
    T_{\mathrm{RMS}} = 
    \frac{\left| \mathrm{RMS}\left( \frac{y_d}{\mathrm{RMS}(x_d)} \right) - \mathrm{RMS}\left( \frac{y_h}{\mathrm{RMS}(x_h)} \right) \right|}{\mathrm{RMS}\left( \frac{y_h}{\mathrm{RMS}(x_h)} \right)}
    \label{eq8}
\end{equation}
where $\mathrm{RMS}(\cdot)$ denotes the root-mean-square operator.

To characterize the evolution of temporal response, a fourth-order autoregressive model was fitted to the normalized roof acceleration response:
\begin{equation}
    y(t) = \sum_{i=1}^{4} a_i y(t - i) + e(t)
    \label{eq9}
\end{equation}
where $a_i$ are the autoregressive coefficients and $e(t)$ is the residual term. The corresponding DSF was defined as:
\begin{equation}
    T_{\mathrm{AR}4} = \left\lVert \mathbf{a}_d - \mathbf{a}_h \right\rVert_{2}
    \label{eq10}
\end{equation}
where $\mathbf{a}_h$ and $\mathbf{a}_d$ denote the AR coefficient vectors corresponding to the healthy and damaged states, respectively.

Finally, an effective stiffness proxy, denoted as $K_{\mathrm{PRX}}$, was introduced based on acceleration-only measurements:
\begin{equation}
    K_{\mathrm{PRX}} = \mathrm{median} \left| \frac{y(t)}{d_y(t) - d_x(t)} \right|
    \label{eq11}
\end{equation}
where $d_x(t)$ and $d_y(t)$ are the corresponding ground and roof displacement responses obtained through frequency-domain double integration. The associated DSF was defined as:
\begin{equation}
    T_{K_{\mathrm{PRX}}} = \frac{K_h - K_d}{K_h}
    \label{eq12}
\end{equation}
where $K_h$ and $K_d$ denote the effective stiffness proxies corresponding to the healthy and damaged states, respectively.

\begin{table}[ht]
    \centering
    \caption{Summary of the selected physics-informed DSFs used for structural condition assessment}
    \label{tab:2}
    \begin{tabular}{llll}
        \toprule
        ID & DSF & Domain & Physical interpretation \\
        \midrule
        1--2 & $F_{\text{peak}}^{(1)},\, F_{\text{peak}}^{(2)}$ & Frequency & Variation of Modal frequency \\
        3--4 & $F_{\text{centroid}}^{(1)},\, F_{\text{centroid}}^{(2)}$ & Frequency & Redistribution of transmissibility spectral energy \\
        5--6 & $F_{\text{MAC}}^{(1)},\, F_{\text{MAC}}^{(2)}$ & Frequency & Shape variation of transmissibility spectrum \\
        7--8 & $F_{\text{area}}^{(1)},\, F_{\text{area}}^{(2)}$ & Frequency & Change in modal-band transmissibility energy \\
        9   & $T_{\text{delay}}$ & Time & Damage-induced response phase-lag variation \\
        10  & $T_{\text{RMS}}$ & Time & Global response amplification variation \\
        11  & $T_{\text{AR4}}$ & Time & Temporal response evolution change \\
        12  & $T_{\text{KPRX}}$ & Time & Effective dynamic stiffness degradation \\
        \bottomrule
    \end{tabular}
\end{table}

\subsubsection{Generic statistical time-series descriptors}
To provide a comparison against manually designed physics-informed DSFs, generic statistical time-series representations were also extracted using the Catch22 feature set \cite{lubba2019catch22}. Catch22 consists of 22 canonical time-series descriptors selected from a large collection of statistical features through extensive benchmarking across diverse classification tasks. These descriptors characterize various statistical and dynamical properties of time-series signals, including distribution characteristics, temporal autocorrelation, fluctuation behavior, entropy-related measures, and nonlinear temporal patterns.

In this study, Catch22 features were extracted directly from the roof acceleration response histories obtained from the healthy-state and post-earthquake white-noise excitation analyses. For each Catch22 descriptor, a relative-change-based DSF was computed as
\begin{equation}
    F_{\mathrm{C22},i}
    =
    |\frac{
    F_{d,i}-F_{h,i}
    }{
    F_{h,i}
    }|,
    \label{eq13}
\end{equation}
where $F_{h,i}$ and $F_{d,i}$ denote the $i$-th Catch22 descriptor extracted from the healthy and damaged roof acceleration responses, respectively. The resulting feature vectors were subsequently used as inputs for ML-based damage classification and transferability evaluation within heterogeneous structural populations.

\subsubsection{Convolution kernel-based representations}
In addition to manually designed and statistical descriptors, convolution kernel-based time-series representations were investigated using the MiniRocket framework \cite{dempster2021minirocket}. MiniRocket is a computationally efficient transformation method derived from the ROCKET family of algorithms, which employs a large number of predefined random convolutional kernels to extract discriminative temporal patterns from raw time-series signals. Consequently, they provide a representative benchmark for evaluating whether generic deep-learning-based time-series representations, without incorporating structural dynamics knowledge or input-output relationships, can achieve transferable post-earthquake damage classification performance across structurally heterogeneous building populations.

In this study, MiniRocket was applied to a four-channel input constructed from the healthy-state and post-earthquake white-noise responses: $\mathbf{X}(t)=\left[x_h(t),y_h(t),x_d(t),y_d(t)\right]$, consisting of the healthy-state ground acceleration, healthy-state roof acceleration, post-earthquake ground acceleration, and post-earthquake roof acceleration under WN signals, respectively. The transformed feature vectors were subsequently used as inputs for ML-based damage classification and transferability evaluation across heterogeneous structural populations. 

\subsubsection{Structural metadata enrichment}
To further investigate transferable damage representations for structurally heterogeneous building populations, structural metadata was additionally incorporated into the feature space. While vibration-response-based features characterize the dynamic behavior of a structure under excitation, they do not explicitly describe the underlying structural configuration. In the context of PBSHM, such structural variability may significantly influence the relationship between measured responses and damage states. Consequently, enriching response-based representations with structural metadata may provide complementary information that improves cross-structure generalization capability.

As summarized in Table \ref{tab:3}, three categories of structural metadata were considered, namely geometrical properties, material parameters, and modal characteristics. The geometrical information included the number of stories, number of bays, story height, bay width, and beam--column section dimensions, which may typically be obtained from design drawings, BIM models, or field measurements. The material-related information included concrete compressive strength, concrete elastic modulus, steel yield strength, steel elastic modulus, and gravity load parameters, which may be predicted from design documentation or estimated through non-destructive evaluation (NDE) techniques. In addition, modal characteristics, including damping ratio and the first two modal frequencies that can be identified from measured vibration data through operational modal analysis techniques, were also incorporated into the feature space.

To systematically investigate the contribution of different metadata categories, an incremental feature-enrichment strategy was adopted. Starting from the response-based feature representations alone, structural metadata was progressively incorporated into the feature space through the following configurations: (1) response features only, (2) response features with geometrical information, (3) response features with geometrical and material information, and (4) response features with geometrical, material, and modal information. This strategy enables systematic evaluation of the relative contribution of different categories of structural metadata to transferable post-earthquake damage classification.

\begin{table}[ht]
    \centering
    \caption{Summary of the structural metadata considered for feature-space enrichment}
    \label{tab:3}
    \begin{tabular}{c c c c l}
        \toprule
        ID & Metadata & Unit & Category & Notes \\
        \midrule
        13 & $n_s$     & /        & Geometric & Number of stories \\
        14 & $n_b$     & /        & Geometric & Number of bays \\
        15 & $H_s$     & m        & Geometric & Story height \\
        16 & $W_b$     & m        & Geometric & Bay width \\
        17 & $H_t$     & m        & Geometric & Total height \\
        18 & $W_t$     & m        & Geometric & Total width \\
        19 & $b_c$     & m        & Geometric & Column section width \\
        20 & $h_c$     & m        & Geometric & Column section depth \\
        21 & $b_b$     & m        & Geometric & Beam section width \\
        22 & $h_b$     & m        & Geometric & Beam section depth \\
        23 & $f_c$     & MPa      & Material  & Concrete strength \\
        24 & $E_c$     & MPa      & Material  & Concrete elastic modulus \\
        25 & $f_y$     & MPa      & Material  & Steel yielding strength \\
        26 & $E_y$     & MPa      & Material  & Steel elastic modulus \\
        27 & $q_D$     & kN/m$^2$ & Material  & Dead load \\
        28 & $q_L$     & kN/m$^2$ & Material  & Live load \\
        29 & $\zeta$  & \%       & Modal     & Damping ratio \\
        30 & $f_1$     & Hz       & Modal     & First-mode frequency \\
        31 & $f_2$     & Hz       & Modal     & Second-mode frequency \\
        \hline
    \end{tabular}
\end{table}

\subsection{ML-based damage classification}
The generated simulation dataset, with labels and the extracted feature representations, was used to train supervised multi-class classifiers for post-earthquake damage state identification. The classification target was the global structural damage state ($DS \in {DS_0, DS_1, DS_2, DS_3}$), defined from the structure-specific drift thresholds described previously. For each simulation sample, the classifier input consisted of one selected feature representation, such as physics-informed DSFs, Catch22 descriptors, MiniRocket representations, or metadata-enriched feature vectors.

To evaluate whether the learned classifiers can generalize across structurally different systems, the dataset was split at the structural-model level rather than at the individual-sample level. Specifically, the structural model ID was used as the grouping variable, such that all earthquake simulations associated with the same structural model were assigned exclusively to either the training or testing set. This group design prevents information leakage and provides a more realistic assessment of cross-structure transferability. In each experiment, 80\% of the structural models were used for training and the remaining 20\% were reserved for testing. To reduce dependence on a particular data partition, the group-based train-test split was repeated using five different random seeds. The identical train-test splits were used for all feature representations to ensure a consistent and fair comparison. Hyperparameter optimization was performed only on the training set using five-fold Stratified Group K-fold cross-validation, again using the model ID as the grouping variable.

Six representative ML classifiers were investigated: Logistic Regression (LR) \cite{hosmer2013applied}, Random Forest (RF) \cite{breiman2001random}, XGBoost \cite{chen2016xgboost}, Support Vector Machine (SVM) \cite{cortes1995support}, Multi-Layer Perceptron (MLP) \cite{rumelhart1986learning}, and k-Nearest Neighbors (KNN) \cite{cover1967nearest}. These models were selected to cover different learning mechanisms, including linear classification, tree-based ensemble learning, kernel-based nonlinear classification, neural-network-based learning, and distance-based classification. LR provides a simple linear baseline for assessing whether the feature space is linearly separable. RF and XGBoost were included to capture nonlinear feature interactions and possible threshold-type decision boundaries. SVM was adopted as a robust kernel-based classifier for medium-dimensional feature spaces, while MLP was used to evaluate the performance of a neural-network-based classifier. KNN was included as a non-parametric baseline that directly reflects local similarity in the feature space. Although more advanced classification algorithms or ensemble strategies, such as stacking, could further combine the strengths of different classifiers, the objective of this study was not to identify the single best-performing ML model. Instead, multiple classifier families were intentionally considered to reduce the risk that the observed conclusions are specific to a particular learning algorithm.

All classifiers were implemented within a unified preprocessing and training pipeline. Continuous input features were standardized using a standard scaler fitted only on the training data. Hyperparameters were optimized using Optuna \cite{akiba2019optuna} with the macro-averaged F1 score ($\text{F1}_\text{macro}$) as the objective function since the generated dataset exhibits class imbalance among damage states (see Figure \ref{DS_definition_distribution}(b)). For each classifier, 100 Optuna trials were performed, and balanced class weights were adopted for classifiers supporting class weighting to mitigate the influence of class imbalance during training.
The search spaces for each classifier in this study are summarized in Table \ref{tab:ml_search_space} in the Appendix. After hyperparameter tuning, each classifier was retrained on the corresponding training set and evaluated on the held-out test set. Model performance was quantified using accuracy, $\text{F1}_\text{macro}$, and row-normalized confusion matrices. 

\subsection{Experimental validation}
\begin{figure}
    \centering
    \includegraphics[width=0.9\linewidth]
    {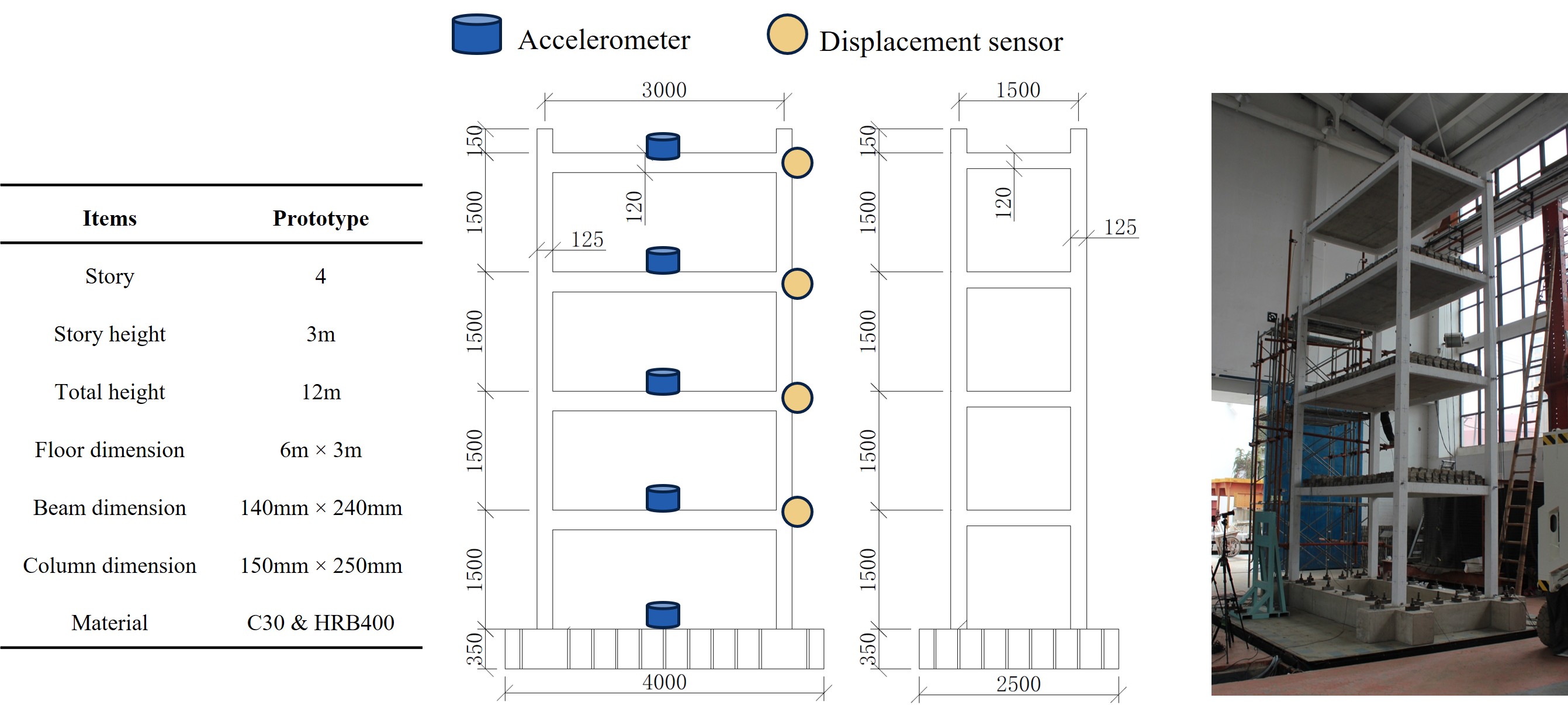}
    \caption{Experimental RC frame structure used for ML-based damage classification validation: structural configuration, specimen dimensions, and test setup \cite{zhang2024post}.}
    \label{exp_setup}
\end{figure}

To further evaluate the transferability of the proposed feature representations under realistic measurement conditions, shake table test data from a half-scale RC frame structure were adopted for external validation \cite{zhang2024post}. As shown in Figure \ref{exp_setup}, the specimen was a cast-in-place four-story RC frame tested under unidirectional seismic excitation. The structure had a story height of 1.5 m and a plan dimension of 1.5 m $\times$ 3.0 m. It was constructed using C30 concrete and HRB400 reinforcing steel, with additional floor masses provided by lead blocks to reproduce gravity effects. During the test, multiple types of sensors, including accelerometers, strain gauges, and displacement sensors, were installed to record the structural responses. 

It is worth noting that the experimental specimen differs substantially from the numerically generated structural population in terms of span configuration, member dimensions, reinforcement detailing, mass distribution, and boundary conditions. Therefore, it represents an out-of-distribution validation case relative to the simulated training population summarized in Table \ref{tab:1}, enabling a more rigorous assessment of whether the proposed feature representations can generalize beyond the numerical parameter space used for model training. 

\begin{figure}[htbp]
    \centering
    \includegraphics[width=0.6\linewidth]
    {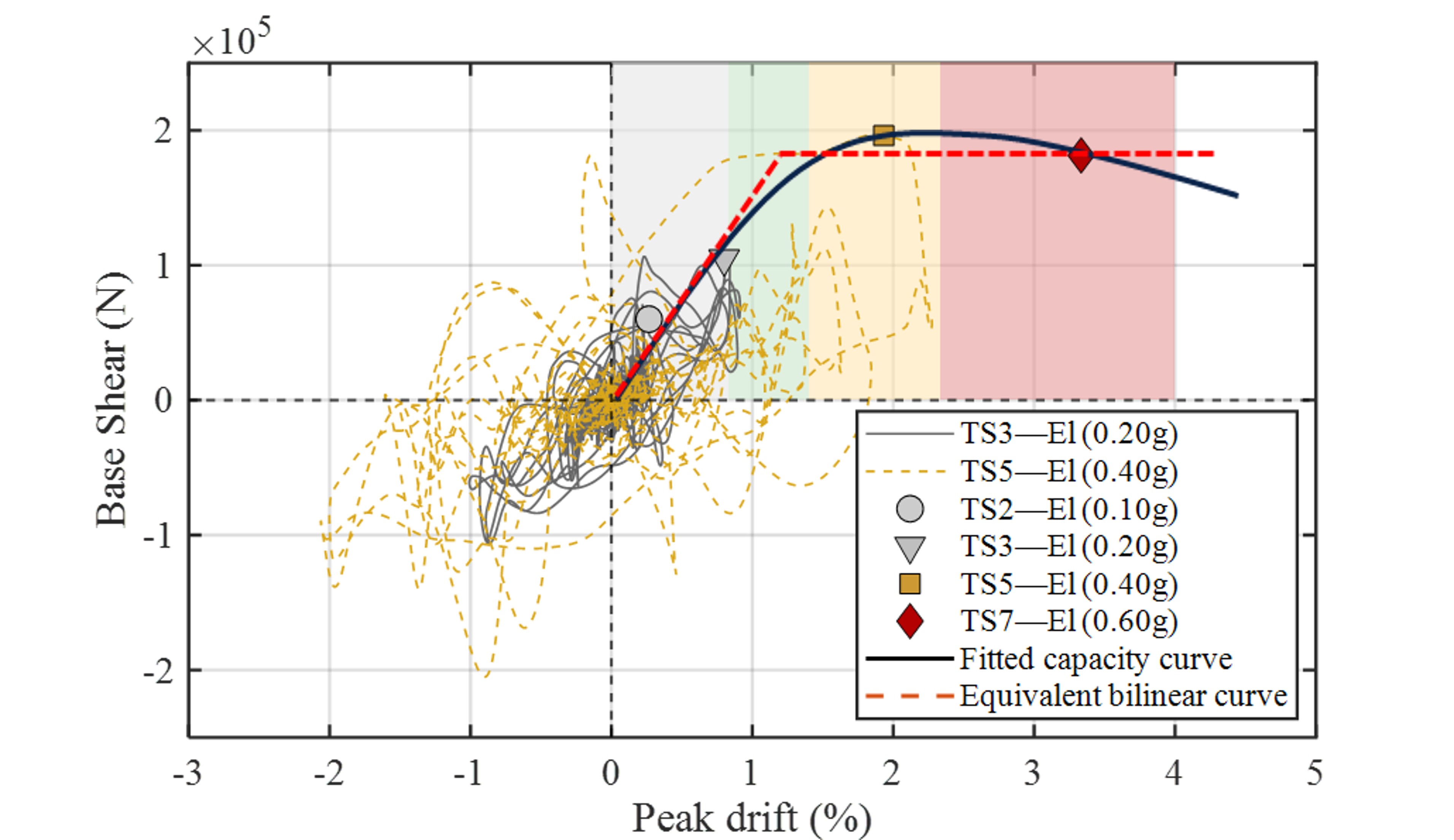}
    \caption{Capacity-curve-based damage state assignment for the experimental shake-table frame structure. The measured base-shear-peak-drift hysteretic responses under different earthquake intensity levels are summarized by representative envelope points, which are used to fit the experimental backbone capacity curve and its equivalent bilinear idealization. The shaded regions indicate the drift-based DS intervals used for experimental label assignment.}
    \label{exp_damage_states}
\end{figure}

The experimental program consisted of a sequence of white-noise (WN) and earthquake excitations with increasing intensity levels, as listed in Table \ref{tab:4}. The El Centro earthquake inputs were used to induce different levels of structural damage; the initial WN test (TS1) was used as the healthy reference state, while the post-earthquake WN tests (TS4, TS6, TS8) were used to characterize the structural dynamic state after different levels of earthquake-induced damage. Particularly, the DS labels used for validation were assigned based on a capacity-curve-based interpretation of the measured responses. As shown in Figure \ref{exp_damage_states}, the hysteretic responses obtained under different earthquake intensity levels were first reduced to representative envelope points. For each earthquake excitation case, the envelope point was defined by the peak drift demand and the corresponding base shear response, thereby characterizing the maximum deformation state reached by the specimen during that excitation. These envelope points were then used to fit the experimental backbone capacity curve. An equivalent bilinear capacity curve was subsequently derived through energy-equivalent idealization, from which drift-based DS thresholds can be defined, as detailed in Figure \ref{DS_definition_distribution}(a). Each earthquake excitation case was then assigned a DS label by comparing its peak drift demand with the corresponding threshold intervals. Visual damage observations were retained only as supplementary references for comparison rather than as the primary labeling criterion. To construct validated sample-label pairs, the WN responses were segmented into multiple time windows, and only the ground and roof acceleration measurements were used for feature extraction, following the same sparse sensing configuration adopted in the numerical simulations. Specifically, each response history in WN tests (TS4, TS6, TS8) was divided into 30-second windows, and all windows inherited the DS label assigned to the corresponding preceding earthquake-induced structural condition, as presented in Table \ref{tab:4}. This procedure generated 57 experimental validation samples in total that were used exclusively for external testing. The classifiers trained on the numerical population dataset were directly applied to the experimental DSF-based and metadata-enriched feature representations without retraining, fine-tuning, or domain adaptation. This validation protocol provides a strict assessment of the simulation-to-experiment transferability of the physics-informed feature framework under realistic measurement uncertainty and model discrepancy.

\begin{table}[htbp]
    \centering
    \caption{Experimental test program and corresponding DS labels used for constructing sample-label pairs}
    \label{tab:4}
    \begin{tabular}{lllll}
        \toprule
        NO. & Input ground motion & Amplitude & DS by test observation \cite{zhang2024post} & DS by capacity curve \\
        \midrule
        TS1 & White Noise & 0.05 g & Healthy reference         \\
        TS2 & El Centro   & 0.10 g & No damage             & $\mathrm{DS}_0$ \\
        TS3 & El Centro   & 0.20 g & Very light damage     & $\mathrm{DS}_0$ \\
        TS4 & White Noise & 0.05 g & --                    & $\mathbf{DS_0}$ \\
        TS5 & El Centro   & 0.40 g & Moderate damage       & $\mathrm{DS}_2$ \\
        TS6 & White Noise & 0.05 g & --                    & $\mathbf{DS_2}$ \\
        TS7 & El Centro   & 0.60 g & Severe damage         & $\mathrm{DS}_3$ \\
        TS8 & White Noise & 0.05 g & --                    & $\mathbf{DS_3}$ \\
        \bottomrule
    \end{tabular}
\end{table}

\vspace{-0.1em}

\subsection{Community-level resilience assessment}
To examine the downstream implications of transferable damage feature representations, the ML-inferred damage states were further integrated into a community-level recovery simulation. This module does not aim to develop a new recovery optimization algorithm, but to evaluate how the timeliness and reliability of monitoring-informed damage tags affect inspection, repair prioritization, functionality restoration, and resilience loss.

\subsubsection{Measurement-enabled automatic damage tagging}
The ML-based classifier provides a probabilistic mechanism for translating monitoring-derived features into building-level damage states. In the community analysis, this classification uncertainty was represented through an SHM-based observation matrix obtained from the row-normalized confusion matrix of the trained model:
\begin{equation}
    \mathbf{P}_{\mathrm{SHM}}(i,j)
    =
    P_{\mathrm{SHM}}(\hat{DS}=j \mid DS=i),
    \label{eq14}
\end{equation}
where \(DS=i\) and \(\hat{DS}=j\) denote the true and estimated damage states, respectively. This matrix was used to simulate automatic damage tagging for monitored buildings without explicitly repeating the full feature extraction and classification process for every building in the virtual community.

\subsubsection{Community recovery simulation}
The community was represented by $N_b$ buildings, each assigned an importance weight $w_i$, a damage-dependent functionality value $q_i(t)$, an inspection duration, and a repair duration determined by its actual damage state. The overall community functionality at time \(t\) was defined as
\begin{equation}
    Q(t)=
    \frac{
    \sum_{i=1}^{N_b} w_i q_i(t)
    }{
    \sum_{i=1}^{N_b} w_i
    }.
    \label{eq15}
\end{equation}
The recovery process was modeled as a sequential inspection-repair process. After the earthquake, buildings first undergo damage assessment, either through manual inspection or SHM-enabled rapid tagging. Repair decisions are made only after the damage assessment stage is completed. For buildings with completed assessment, repair priority was determined based on the estimated damage state:
\begin{equation}
    \mathrm{Priority}_i = \frac{\Delta Q_i}{t_{\mathrm{repair},i}}
    \label{eq16}
\end{equation}
where \(\Delta Q_i\) denotes the expected increase in community functionality if building \(i\) is restored, and \(t_{\mathrm{repair},i}\) is the estimated repair duration based on the inferred damage state. Available repair crews were assigned to buildings with the highest priority values. To account for imperfect damage assessment, the recovery simulation distinguished between ordinary damage state mismatches and missed detections. The actual repair duration was increased by a penalty factor for a mismatch \(\hat{DS}\neq DS\) to represent inefficient repair planning. A missed detection was defined as \(\hat{DS}=0\) while \(DS>0\); such buildings were initially excluded from the main repair queue and were repaired later with a larger penalty to represent delayed discovery and additional coordination cost. Conversely, if an undamaged building was incorrectly assessed as damaged, no physical repair was required, but a mobility-time penalty was imposed to represent unnecessary crew dispatch. 

The resilience loss was quantified as the accumulated functionality loss over the recovery horizon:
\begin{equation}
    LoR = \int_{0}^{T}\left[1-Q(t)\right]dt
    \label{eq17}
\end{equation}
where a smaller \(LoR\) indicates faster and more efficient recovery.

\subsubsection{Virtual community setup and information scenarios}
A virtual community composed of 80 RC frame buildings was considered to illustrate the resilience implications of monitoring-informed damage classification. The community includes buildings with different functional roles, including critical facilities and ordinary residential buildings. Each building was assigned an importance weight ($w$), a functionality value ($q$), an inspection duration ($t_{inspect}$), and a repair duration ($t_{repair}$) according to its building type and damage state. The main parameters adopted for the community recovery simulation are summarized in Table \ref{tab:5}. The initial post-earthquake damage condition of each building was sampled from the damage state distribution obtained from the numerical earthquake simulations under the selected seismic intensity scenario. 

Two information scenarios were compared. In the baseline scenario without SHM, all buildings require conventional manual inspection before repair decisions can be made. The inspection outcome is uncertain and was sampled from the following inspection-based observation matrix:
\begin{equation}
\mathbf{P}_{\mathrm{INS}}(\hat{DS}\mid DS)=
\begin{bmatrix}
0.80 & 0.20 & 0.00 & 0.00 \\
0.15 & 0.65 & 0.17 & 0.03 \\
0.03 & 0.17 & 0.65 & 0.15 \\
0.00 & 0.05 & 0.15 & 0.80
\end{bmatrix},
\label{eq18}
\end{equation}
where each row corresponds to the true damage state and each column corresponds to the estimated damage state. This matrix represents moderate uncertainty in rapid visual inspection, with misclassification mainly assumed to occur between adjacent damage states.

In the SHM-informed scenario, monitored buildings receive automatic damage tags from the ML-based damage classifier, with classification uncertainty governed by \(\mathbf{P}_{\mathrm{SHM}}\). The tagging time for SHM-monitored buildings was set to one-tenth of the corresponding manual inspection duration for the same building type and damage state, representing the accelerated information acquisition enabled by automated monitoring. After damage assessment is completed, buildings in both scenarios follow the same repair priority rule, repair crew constraints, and damage-dependent repair assumptions. Therefore by comparing the resulting functionality trajectories and resilience loss values, the effect of SHM-enabled damage tagging on community-level recovery performance can be quantified.

\begin{table}[htbp]
\centering
\caption{Parameters adopted for the virtual community recovery simulation. 
The entries in the DS-dependent columns follow the order of $[DS_0, DS_1, DS_2, DS_3]$. Additional recovery assumptions: mismatch penalty factor = 1.4, missed-detection penalty factor = 2.0, and mobility-time penalty for unnecessary dispatch = 0.5 days.} 
\label{tab:5}
\renewcommand{\arraystretch}{1.15}
\setlength{\tabcolsep}{4pt}
\small
\begin{tabular}{lccccccc}
\toprule
Type & $N$ & $w$ & $\mathbf{q}$ & $\mathbf{t}_{\mathrm{inspect}}$ & $\mathbf{t}_{\mathrm{repair}}$ & $N_{\mathrm{inspect}}$ & $N_{\mathrm{repair}}$ \\
\midrule
Hospital 
& 8 & 5 & [1.0, 0.7, 0.3, 0.0]& [0.20, 0.30, 0.40, 0.50]& [0, 5, 15, 35]& \multirow{2}{*}{3}& \multirow{2}{*}{5}  \\

Residential 
& 72 & 1 & [1.0, 0.8, 0.4, 0.0] & [0.20, 0.25, 0.30, 0.35] & [0, 5, 15, 35] &  &  \\
\hline
\end{tabular}
\end{table}

\section{Results and Discussions}
\subsection{Individual DSF performance}
Before evaluating multi-feature classification models, the damage-discriminative capability of individual indicators was first examined using an indicator-based fragility analysis. The purpose was to assess whether a single scalar indicator can act as an observable proxy for structural damage state and to reveal the limitations of single-feature-based classification across heterogeneous structural systems. For each scalar indicator \(x\), including conventional intensity measures (IMs, e.g., PGA), engineering demand parameters (EDPs), and physics-informed DSFs, lognormal vulnerability curves were fitted on the training set to estimate the exceedance probabilities:
\begin{equation}
    P(DS \geq k \mid x)
    = \Phi \left(\frac{\ln x-\ln \theta_k}{\beta_k} \right), \quad k=1,2,3
    \label{eq19}
\end{equation}
where \(\Phi(\cdot)\) denotes the standard normal cumulative distribution function, while \(\theta_k\) and \(\beta_k\) are the median and logarithmic dispersion of the vulnerability curve associated with the exceedance threshold \(DS\geq k\), respectively. The parameters were estimated by maximum likelihood using only the training structures, while the reported accuracy and \(F1_{\mathrm{macro}}\) were evaluated on unseen test structures. This setting is consistent with the group-wise transferability evaluation adopted throughout this study (seen in section 2.4). The fitted exceedance probabilities were then converted into discrete DS probabilities as follows:
\begin{equation}
\begin{aligned}
P(DS=0\mid x) &= 1-P(DS\geq1\mid x),\\
P(DS=1\mid x) &= P(DS\geq1\mid x)-P(DS\geq2\mid x),\\
P(DS=2\mid x) &= P(DS\geq2\mid x)-P(DS\geq3\mid x),\\
P(DS=3\mid x) &= P(DS\geq3\mid x).
\end{aligned}
\label{eq20}
\end{equation}
For each test sample from the held-out test structures, the final damage state prediction was assigned according to the maximum discrete DS probability.

\begin{figure}[htbp]
    \centering
    \includegraphics[width=1\linewidth]
    {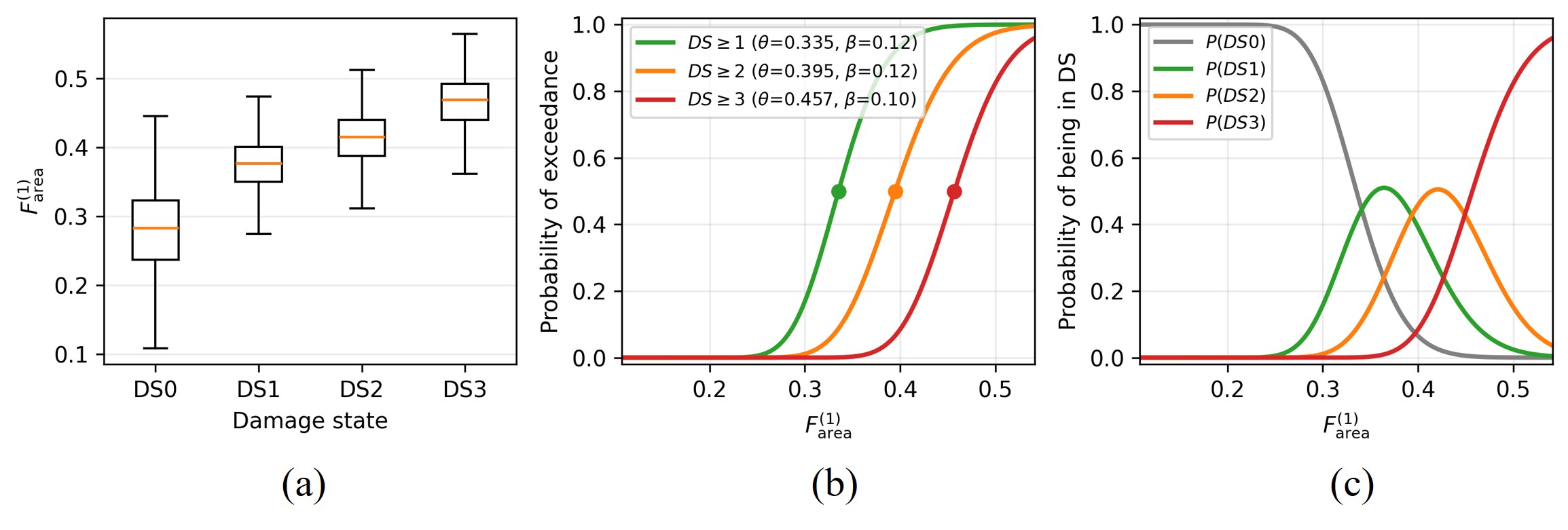}
    \caption{Representative individual-DSF analysis using $F_{\mathrm{area}}^{(1)}$: (a) distribution of DSF values across damage states; 
    (b) fitted lognormal fragility curves \(P(DS\geq k\mid x)\); (c) corresponding discrete DS probability curves \(P(DS=k\mid x)\).}
    \label{single_dsf_fragility}
\end{figure}

Figure \ref{single_dsf_fragility} illustrates this procedure using the best-performing individual DSFs, $F_{\mathrm{area}}^{(1)}$. Figure \ref{single_dsf_fragility}(a) shows that \(F_{\mathrm{area}}^{(1)}\) generally increases with damage severity, indicating that the change in the first modal-band transmissibility energy is sensitive to earthquake-induced structural degradation. The fitted vulnerability curves in Figure \ref{single_dsf_fragility}(b) exhibit monotonic exceedance probabilities for \(DS\geq1\), \(DS\geq2\), and \(DS\geq3\), which are then transformed into discrete DS probabilities according to Eq.(\ref{eq20}). As presented in Figure \ref{single_dsf_fragility}(c), although the most probable damage state shifts progressively from \(DS_0\) to \(DS_3\) as the feature value increases, overlapping probability regions remain between adjacent damage states, indicating the intrinsic limitation of single-feature-based damage classification.

Table \ref{tab:6} summarizes the test-set performance of individual indicators over multiple random group-wise train--test splits. Among the non-DSF indicators, PGA showed relatively poor performance with an average accuracy of 0.544 and an average \(\mathrm{F1}_{\mathrm{macro}}\) of 0.403, indicating that earthquake intensity alone cannot fully capture the damage states of structurally heterogeneous buildings. In comparison, the maximum inter-story drift ratio $\theta$ achieved the highest classification performance, with an average accuracy of 0.780 and an average \(\mathrm{F1}_{\mathrm{macro}}\) of 0.766. This result is expected because the DS labels were defined based on drift-related structural demand. Nevertheless, direct measurement of inter-story drift is generally difficult in practical monitoring applications, especially for large building portfolios, where reliable displacement measurements or dense sensing configurations are rarely available. Therefore, $\theta$ mainly serves as a reference indicator rather than a readily observable feature for scalable PBSHM. In this context, the acceleration-based DSFs introduced in section 2.3.1 offer a remedy. As compared, the first-mode frequency-domain DSFs generally outperformed, suggesting that damage-induced changes in the first-mode transmissibility band provide stronger and more stable information for DS inference in PBSHM. Nevertheless, the performance of all individual DSFs remains lower than that of $\theta$, and several indicators show limited standalone discriminative capacity. These results demonstrate that individual DSFs capture only partial aspects of earthquake-induced structural degradation, thereby motivating the subsequent fusion of multiple DSFs through ML-based classifiers to improve performance.

\begin{table}[ht]
    \centering
    \caption{Test-set performance of individual IMs, EDPs, and physics-informed DSFs using vulnerability-based single-indicator DS inference. Mean and standard deviation are computed over multiple random group-wise train--test splits.}
    \label{tab:6}
    \setlength{\tabcolsep}{11pt}   
    \begin{tabular}{llcccc}
        \toprule
        \multirow{2}{*}{Category} & \multirow{2}{*}{Feature} & \multicolumn{2}{c}{Accuracy} & \multicolumn{2}{c}{$\mathrm{F1}_\mathrm{macro}$} \\
        \cmidrule(r){3-4} \cmidrule(r){5-6} 
         &  & Mean & Std & Mean & Std \\
        \midrule
        IM  & PGA              & 0.544 & 0.009 & 0.403 & 0.003 \\
        EDP & MRD              & 0.737 & 0.013 & 0.718 & 0.016 \\
        EDP & RRD              & 0.513 & 0.013 & 0.461 & 0.011 \\
        EDP & $\theta$        & \textbf{0.780} & \textbf{0.015} & \textbf{0.766} & \textbf{0.017} \\
        DSF & $F_{\text{centroid}}^{(1)}$ & 0.576 & 0.042 & 0.471 & 0.030 \\
        DSF & $F_{\text{MAC}}^{(1)}$      & 0.483 & 0.036 & 0.328 & 0.021 \\
        DSF & $F_{\text{peak}}^{(1)}$     & 0.630 & 0.042 & 0.608 & 0.043 \\
        DSF & $F_{\text{area}}^{(1)}$ & \textbf{0.650} & \textbf{0.024} & \textbf{0.625} & \textbf{0.020} \\
        DSF & $F_{\text{centroid}}^{(2)}$ & 0.354 & 0.044 & 0.214 & 0.029 \\
        DSF & $F_{\text{MAC}}^{(2)}$      & 0.318 & 0.016 & 0.189 & 0.002 \\
        DSF & $F_{\text{peak}}^{(2)}$     & 0.566 & 0.020 & 0.439 & 0.055 \\
        DSF & $F_{\text{area}}^{(2)}$     & 0.329 & 0.024 & 0.193 & 0.015 \\
        DSF & $T_{\text{delay}}$          & 0.313 & 0.011 & 0.176 & 0.006 \\
        DSF & $T_{\text{RMS}}$            & 0.401 & 0.008 & 0.203 & 0.009 \\
        DSF & $T_{\text{AR4}}$            & 0.430 & 0.022 & 0.278 & 0.027 \\
        DSF & $T_{\text{KPRX}}$           & 0.243 & 0.003 & 0.098 & 0.001 \\
        \bottomrule
    \end{tabular}
\end{table}

\subsection{ML-based DSF fusion performance}
\begin{figure}[htbp]
    \centering
    \includegraphics[width=1\linewidth]
    {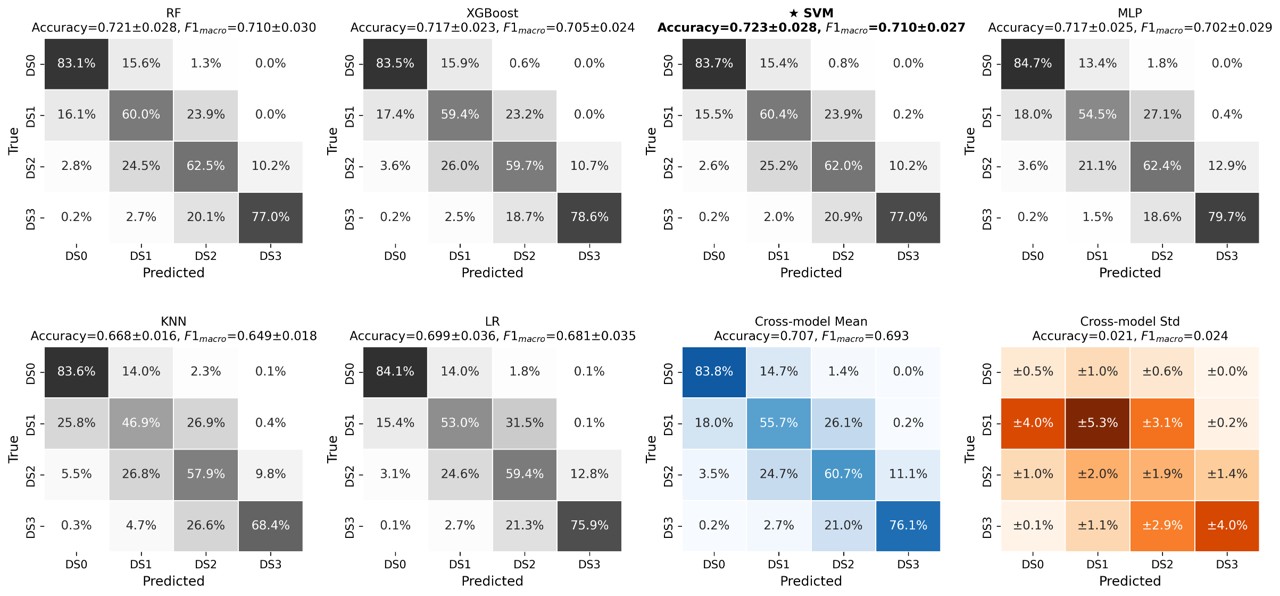}
    \caption{Row-normalized confusion matrices of the investigated ML classifiers using the fused physics-informed DSFs representation over multiple random group-wise train–test splits. The cross-model mean and standard deviation summarize the average and variability of classification behavior across the six classifiers.}
    \label{ml_dsf_confusion}
\end{figure}

Figure \ref{ml_dsf_confusion} provides the test performance of six classifiers trained using the fused 12-dimensional physics-informed DSFs listed in Table \ref{tab:2}. Compared with the best individual DSF, \(F_{\mathrm{area}}^{(1)}\), which achieved an average \(\mathrm{F1}_{\mathrm{macro}}\) of 0.625, the fused DSF representation led to a clear improvement. The overall mean performance across the six classifiers reached an accuracy of 0.707 and an average \(\mathrm{F1}_{\mathrm{macro}}\) of 0.693, confirming that different DSFs provide complementary damage-related information and that their combined representation is more effective for distinguishing structural damage states across heterogeneous building populations.

In addition, the performance differences among the investigated classifiers were relatively small, as reflected by the limited cross-model standard deviations. SVM achieved the highest average performance, with an accuracy of 0.723 and an \(\mathrm{F1}_{\mathrm{macro}}\) of 0.710, while RF, XGBoost, MLP, and LR showed comparable results. This indicates that the improvement is not tied to a specific classifier but mainly arises from the fused physics-informed feature representation. KNN showed relatively lower performance, suggesting that simple local similarity in the DSF space is less effective for separating damage states across heterogeneous structures.

Furthermore, the row-normalized confusion matrices reveal that $DS_0$ and $DS_3$ were identified more reliably than the intermediate damage states. The cross-model mean correct classification rates were 83.8\% for $DS_0$ and 76.1\% for $DS_3$, whereas $DS_1$ and $DS_2$ showed lower diagonal rates of 55.7\% and 60.7\%, respectively. Most misclassifications occurred between adjacent damage states, especially between $DS_1$ and $DS_2$. This pattern is consistent with the gradual transition of structural damage severity and the overlap observed in the single-DSF probability curves. Overall, ML-based fusion of physics-informed DSFs provides a more robust damage representation than individual indicators, while the separation of moderate damage states remains challenging.

\subsection{Comparison between feature representations}
Table \ref{tab:7} compares the classification performance of the three representation paradigms, i.e., fused DSFs, Catch22 descriptors, and MiniRocket representations, across the six investigated ML classifiers under the same group-wise training and testing protocol. The comparison shows that the physics-informed DSF representation consistently outperformed the other two across all the investigated classifiers. Specifically, Catch22 descriptors led to average reductions of 24.2\% in accuracy and 29.6\% in \(\mathrm{F1}_{\mathrm{macro}}\), while MiniRocket showed reductions of 20.1\% and 25.5\%, respectively. 

This performance gap indicates that generic statistical time-series Catch22 descriptors and convolution kernel-based MiniRocket representations, although effective in many general classification tasks, are less transferable for post-earthquake damage state classification across structurally heterogeneous systems. Neither of them explicitly encodes structural dynamics mechanisms such as transmissibility variation, modal-band energy redistribution, or effective stiffness degradation. As a result, they may be more sensitive to structural variability and excitation-dependent signal patterns. In contrast, the physics-informed DSFs were manually designed to compare healthy and post-earthquake dynamic responses through physically interpretable input--output relationships. This enables them to better capture damage-induced changes while reducing dependence on structure-specific signal patterns. The consistent superiority of DSFs across different classifiers suggests that the improved performance mainly arises from the representation itself rather than from a particular ML model. These results support the use of physics-informed feature engineering as a more transferable representation strategy for PBSHM-oriented post-earthquake damage classification.

\begin{table}[ht]
    \centering
    \caption{Comparison of classification performance using physics-informed DSFs, Catch22 descriptors, and MiniRocket representations. Mean and standard deviation are computed over multiple random group-wise train--test splits. The percentage changes are calculated relative to the DSF representation.}
    \label{tab:7}
    \setlength{\tabcolsep}{4pt}      
    \begin{tabular}{lcccccc}
        \toprule
        \multirow{2}{*}{ML models} & \multicolumn{2}{c}{DSF} & \multicolumn{2}{c}{Catch22} & \multicolumn{2}{c}{MiniRocket} \\
        \cmidrule(r){2-3} \cmidrule(r){4-5} \cmidrule(r){6-7}
         & Accuracy & $\mathrm{F1}_\mathrm{macro}$ & Accuracy & $\mathrm{F1}_\mathrm{macro}$ & Accuracy & $\mathrm{F1}_\mathrm{macro}$ \\
        \midrule
        RF     & $0.721\pm0.028$ & $0.710\pm0.030$ & $0.540\pm0.032$ & $0.487\pm0.042$ & $0.571\pm0.033$ & $0.500\pm0.037$ \\
        XGBoost& $0.717\pm0.023$ & $0.705\pm0.024$ & $0.552\pm0.030$ & $0.508\pm0.034$ & $0.558\pm0.032$ & $0.512\pm0.037$ \\
        SVM    & $0.723\pm0.028$ & $0.710\pm0.027$ & $0.558\pm0.031$ & $0.510\pm0.032$ & $0.570\pm0.040$ & $0.534\pm0.043$ \\
        MLP    & $0.717\pm0.025$ & $0.702\pm0.029$ & $0.515\pm0.022$ & $0.475\pm0.029$ & $0.570\pm0.026$ & $0.517\pm0.038$ \\
        KNN    & $0.668\pm0.016$ & $0.649\pm0.018$ & $0.491\pm0.013$ & $0.437\pm0.019$ & $0.597\pm0.033$ & $0.558\pm0.032$ \\
        LR     & $0.699\pm0.036$ & $0.681\pm0.035$ & $0.557\pm0.043$ & $0.510\pm0.039$ & $0.524\pm0.051$ & $0.472\pm0.055$ \\
        Cross-model   & $0.707\pm0.021$ & $0.693\pm0.024$ & $0.536\pm0.027$ & $0.488\pm0.029$ & $0.565\pm0.022$ & $0.516\pm0.028$ \\
        Change & /               & /               & $-24.2\%$       & $-29.6\%$       & $-20.1\%$       & $-25.5\%$       \\
        \bottomrule
    \end{tabular}
\end{table}

\subsection{Effect of structural metadata enrichment}
The effect of structural metadata enrichment was further investigated by progressively augmenting the physics-informed DSF representation with geometrical information (GeoI), material information (MatI), and modal properties (ModI) given in Table \ref{tab:3}. Table \ref{tab:8} summarizes the classification performance of the three metadata-enriched configurations: DSF+GeoI, DSF+GeoI+MatI, and DSF+GeoI+MatI+ModI, while the DSF-only representation reported in the previous section is used as the baseline for calculating the relative performance changes. As shown, adding geometrical information led to a moderate improvement over the DSF-only baseline, increasing the mean accuracy and \(\mathrm{F1}_{\mathrm{macro}}\) score by 4.8\% and 4.3\%, respectively. A more pronounced improvement was obtained after further incorporating material information, with the DSF+GeoI+MatI representation increasing the mean accuracy and \(\mathrm{F1}_{\mathrm{macro}}\) score by 10.6\% and 10.7\%, respectively, indicating the complementary role of geometry and material-related parameters in providing useful contextual information for interpreting vibration-derived DSFs across heterogeneous buildings. In contrast, the additional inclusion of modal properties led to only marginal improvement, with the mean \(\mathrm{F1}_{\mathrm{macro}}\) score increasing slightly from 0.767 to 0.768. This limited gain may be attributed to the fact that modal information is partly redundant with the physics-informed DSFs, which already encode modal-band transmissibility changes, peak-frequency shifts, and spectral-energy redistribution.

\begin{figure}[htbp]
    \centering
    \includegraphics[width=0.9\linewidth]
    {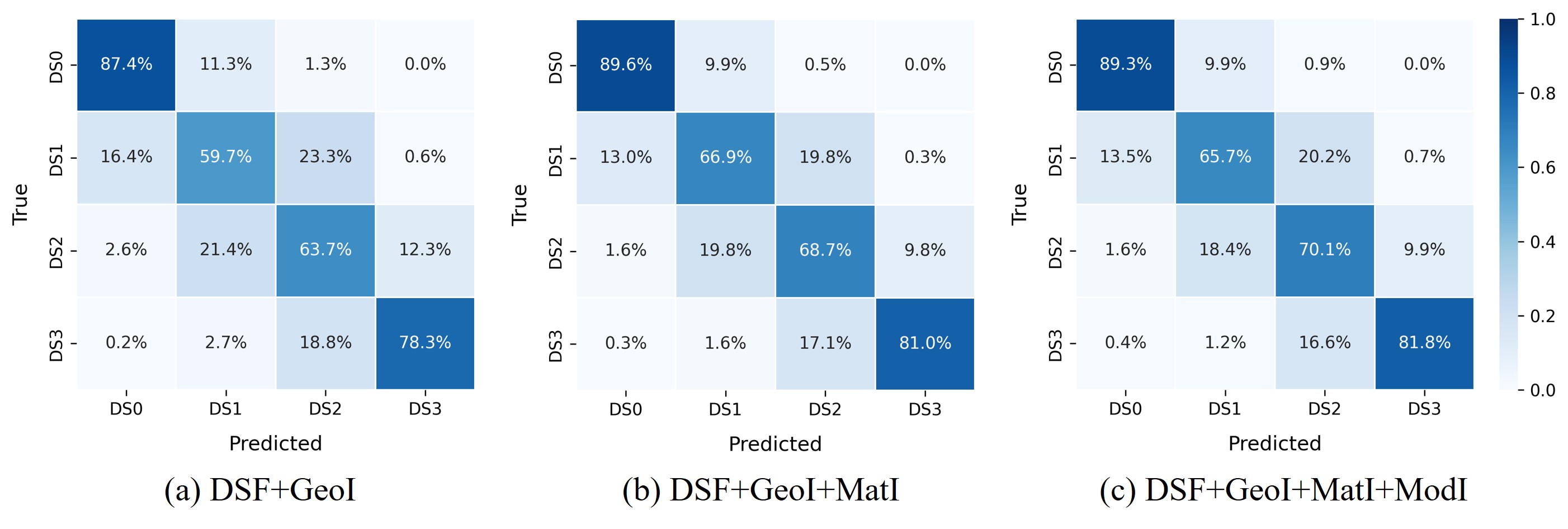}
    \caption{Row-normalized confusion matrices for DSF-based damage classification with structural metadata enrichment.}
    \label{metadata_confusion}
\end{figure}

The cross-model mean confusion matrices provided in Figure \ref{metadata_confusion} further show that structural metadata enrichment mainly improves the classification of intermediate damage states. Compared with the DSF-only baseline in Figure \ref{ml_dsf_confusion}, the diagonal rates for DS1 and DS2 increased from 55.7\% and 60.7\% to 65.7\% and 70.1\% after full metadata enrichment, respectively. Misclassification between adjacent states was reduced, although DS1-DS2 confusion remained the dominant error mode. 

\begin{table}[ht]
    \centering
    \caption{Classification performance of DSF-based representations with progressive structural metadata enrichment. Mean and standard deviation are computed over multiple random group-wise train--test splits. Percentage changes are calculated relative to the DSF-only representation reported in Table \ref{tab:7}.}
    \label{tab:8}
    \begin{tabular}{lcccccc}
        \toprule
        \multirow{2}{*}{ML models} & \multicolumn{2}{c}{DSF+Geol} & \multicolumn{2}{c}{DSF+Geol+Matl} & \multicolumn{2}{c}{DSF+Geol+Matl+Modl} \\
        \cmidrule(r){2-3} \cmidrule(r){4-5} \cmidrule(r){6-7}
        & Accuracy & $\mathrm{F1}_\mathrm{macro}$ & Accuracy & $\mathrm{F1}_\mathrm{macro}$ & Accuracy & $\mathrm{F1}_\mathrm{macro}$ \\
        \midrule
        RF    & 0.744$\pm$0.027 & 0.729$\pm$0.026 & 0.751$\pm$0.030 & 0.735$\pm$0.031 & 0.738$\pm$0.017 & 0.723$\pm$0.015 \\
        XGBoost & 0.756$\pm$0.023 & 0.740$\pm$0.021 & 0.784$\pm$0.027 & 0.768$\pm$0.029 & 0.791$\pm$0.025 & 0.777$\pm$0.024 \\
        SVM   & 0.766$\pm$0.026 & 0.749$\pm$0.027 & 0.832$\pm$0.017 & 0.818$\pm$0.017 & 0.835$\pm$0.016 & 0.822$\pm$0.017 \\
        MLP   & 0.747$\pm$0.027 & 0.724$\pm$0.027 & 0.824$\pm$0.020 & 0.809$\pm$0.022 & 0.827$\pm$0.025 & 0.813$\pm$0.028 \\
        KNN   & 0.677$\pm$0.024 & 0.657$\pm$0.026 & 0.701$\pm$0.018 & 0.684$\pm$0.019 & 0.699$\pm$0.029 & 0.682$\pm$0.031 \\
        LR    & 0.755$\pm$0.024 & 0.738$\pm$0.027 & 0.802$\pm$0.020 & 0.786$\pm$0.021 & 0.804$\pm$0.023 & 0.789$\pm$0.026 \\
        Cross-model  & 0.741$\pm$0.032 & 0.723$\pm$0.033 & 0.782$\pm$0.049 & 0.767$\pm$0.051 & 0.782$\pm$0.053 & 0.768$\pm$0.055 \\
        Change & $+4.8\%$  &  $+4.3\%$  & $+10.6\%$  & $+10.7\%$  & $+10.6\%$  &  $+10.8\%$  \\
        \bottomrule
    \end{tabular}
\end{table}

\vspace{-0.1em}

\begin{figure}[htbp]
    \centering
    \includegraphics[width=1\linewidth]
    {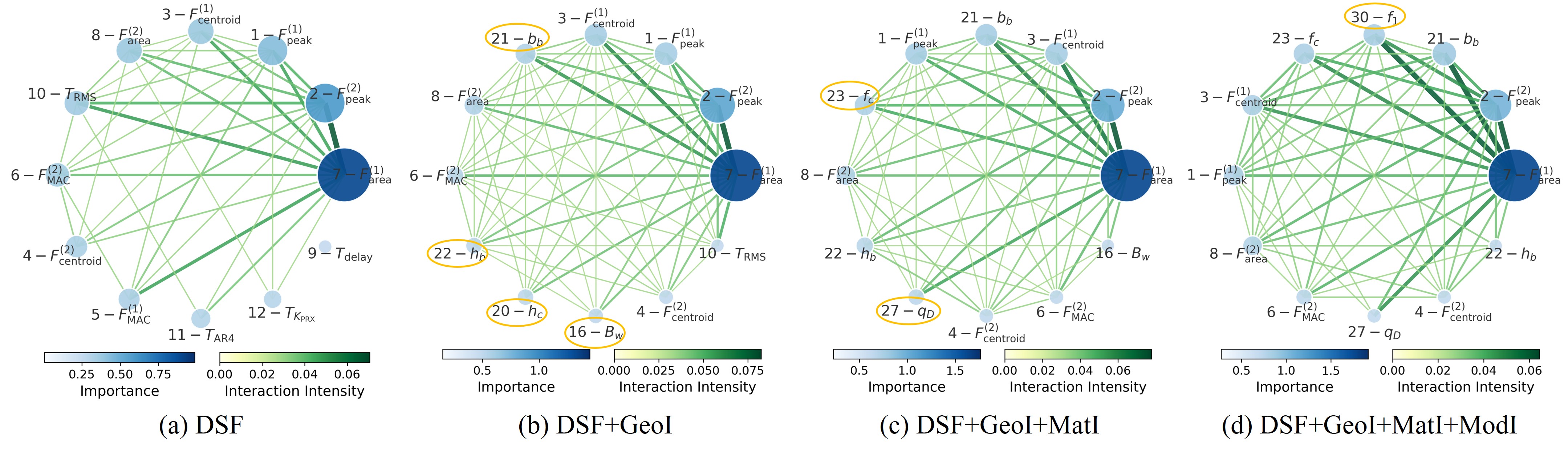}
    \caption{XGBoost-based feature importance and interaction analysis for DSF-based representations with progressive metadata enrichment. Node size indicates feature importance, and edge thickness indicates interaction intensity. Highlighted nodes denote newly introduced structural metadata at each enrichment stage, as summarized in Table \ref{tab:3}.}
    \label{metadata_interaction}
\end{figure}

To further investigate the contributions of physics-informed DSFs and structural metadata, a SHAP-based feature importance and interaction analysis was conducted using XGBoost as a representative tree-based classifier \cite{lundberg2020local}, because it achieved competitive classification performance and enabled convenient evaluation of nonlinear feature contributions and pairwise interactions. As shown in Figure \ref{metadata_interaction}, only the 12 most important features are displayed for readability; node size denotes the mean absolute SHAP value of each feature, and edge thickness represents the SHAP-based interaction intensity between feature pairs. 

In the DSF-only case, \(F_{\mathrm{area}}^{(1)}\) appears as the dominant feature, consistent with its best standalone performance in Table \ref{tab:6}. More generally, the SHAP ranking is dominated by frequency-domain descriptors, whereas time-domain features exhibit comparatively lower individual importance. This suggests that frequency-domain DSFs provide the primary discriminative information for damage characterization. More importantly, the interaction network reveals that $F_{\mathrm{area}}^{(1)}$ does not act independently but interacts strongly with other features, including $F_{\mathrm{peak}}^{(1)}$, $F_{\mathrm{centroid}}^{(1)}$, second-mode descriptors, and $T_{\mathrm{RMS}}$. Although the individual importance of time-domain features is generally lower, their strong interactions with the dominant frequency-domain descriptors suggest that they enhance classification by providing complementary temporal information rather than acting as standalone predictors. Together, these observations indicate that the proposed ML-based feature fusion benefits from the synergy among multiple physically interpretable descriptors rather than relying on any single dominant DSF.

After metadata enrichment, intrinsic structural properties such as member cross-section dimensions, concrete strength, load-related parameters, and the first-mode frequency appear among the important features and interact with the dominant DSFs. This indicates that structural metadata mainly acts as contextual information, helping the classifier interpret similar vibration-derived DSF values under different structural configurations, material capacities, and dynamic characteristics. Overall, the SHAP analysis supports the interpretation that metadata enrichment improves PBSHM transferability by conditioning the DSF-DS relationship rather than simply increasing feature dimensionality.

\subsection{Experimental validation performance}
\begin{figure}[htbp]
    \centering
    \includegraphics[width=1.0\linewidth]
    {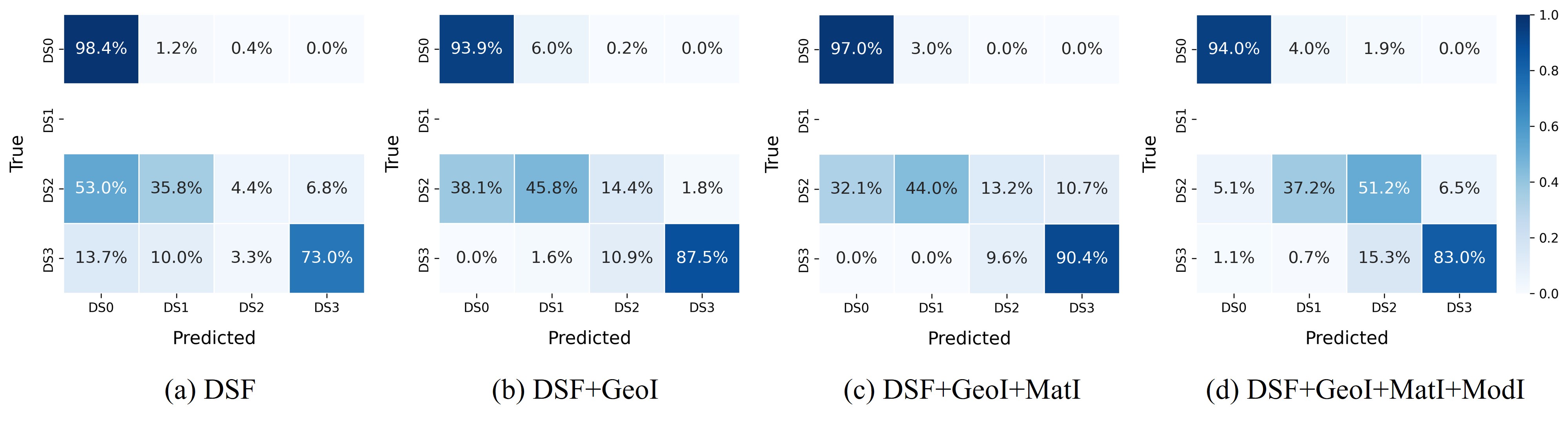}
    \caption{Experimental validation results using simulation-trained classifiers: row-normalized confusion matrices for DSF, DSF+GeoI, DSF+GeoI+MatI, and DSF+GeoI+MatI+Fs representations.}
    \label{experiment_confusion}
\end{figure}

The simulation-trained classifiers were further evaluated on the experimental shaking-table samples described in Section 2.5. No experimental samples were used for model training, hyperparameter tuning, or domain adaptation. Table \ref{tab:9} summarizes the classification performance of the DSF-based representations with progressive structural metadata enrichment. The DSF-only representation showed limited direct transferability to the experimental data, with a mean accuracy of 0.587 and a mean \(\mathrm{F1}_{\mathrm{macro}}\) of 0.404. Adding geometrical and material information improved the results, while the most substantial gain was obtained after incorporating modal property information. Specifically, the DSF+GeoI+MatI+ModI representation achieved a mean accuracy of 0.761 and a mean \(F1\) score of 0.583, corresponding to improvements of 29.6\% and 44.3\% over the DSF-only case. 
The confusion matrices in Figure \ref{experiment_confusion} further indicate that the main challenge lies in identifying the moderate damage condition. The DSF-only representation correctly classifies DS0 and DS3 with high rates, but most DS2 samples are confused with lower damage states. With metadata enrichment, especially after adding modal information, the DS2 diagonal rate increases to 51.2\%, while high recognition rates are maintained for DS0 and DS3, revealing that metadata enrichment mainly reduces ambiguity in the intermediate damage range, where vibration-based features alone are more difficult to interpret.

\begin{table}[ht]
    \centering
    \caption{Experimental validation performance of simulation-trained classifiers using DSF-based representations with progressive structural metadata enrichment.}
    \label{tab:9}
    \resizebox{\textwidth}{!}{%
    \begin{tabular}{lcccccccc}
        \toprule
        \multirow{2}{*}{ML models} & \multicolumn{2}{c}{DSF} & \multicolumn{2}{c}{DSF+Geol} & \multicolumn{2}{c}{DSF+Geol+Matl} & \multicolumn{2}{c}{DSF+Geol+Matl+ModI} \\
        \cmidrule(r){2-3} \cmidrule(r){4-5} \cmidrule(r){6-7}\cmidrule(r){8-9}
         & Accuracy & $\mathrm{F1}_\mathrm{macro}$ & Accuracy & $\mathrm{F1}_\mathrm{macro}$ & Accuracy & $\mathrm{F1}_\mathrm{macro}$ & Accuracy & $\mathrm{F1}_\mathrm{macro}$ \\
        \midrule
        RF     & 0.667$\pm$0.001 & 0.442$\pm$0.010 & 0.667$\pm$0.001 & 0.462$\pm$0.010 & 0.667$\pm$0.002 & 0.459$\pm$0.010 & 0.730$\pm$0.044 & 0.547$\pm$0.059 \\
        XGBoost& 0.702$\pm$0.018 & 0.493$\pm$0.029 & 0.751$\pm$0.040 & 0.582$\pm$0.051 & 0.670$\pm$0.008 & 0.481$\pm$0.017 & 0.768$\pm$0.096 & 0.593$\pm$0.099 \\
        SVM    & 0.449$\pm$0.160 & 0.287$\pm$0.142 & 0.628$\pm$0.146 & 0.481$\pm$0.112 & 0.677$\pm$0.188 & 0.548$\pm$0.247 & 0.681$\pm$0.130 & 0.520$\pm$0.115 \\
        MLP    & 0.339$\pm$0.081 & 0.215$\pm$0.042 & 0.568$\pm$0.076 & 0.422$\pm$0.074 & 0.505$\pm$0.209 & 0.370$\pm$0.175 & 0.604$\pm$0.182 & 0.452$\pm$0.149 \\
        KNN    & 0.719$\pm$0.033 & 0.539$\pm$0.039 & 0.674$\pm$0.016 & 0.570$\pm$0.032 & 0.684$\pm$0.025 & 0.526$\pm$0.030 & 0.877$\pm$0.058 & 0.686$\pm$0.034 \\
        LR     & 0.646$\pm$0.023 & 0.448$\pm$0.011 & 0.628$\pm$0.044 & 0.478$\pm$0.050 & 0.807$\pm$0.104 & 0.682$\pm$0.198 & 0.905$\pm$0.079 & 0.700$\pm$0.046 \\
        Mean   & 0.587$\pm$0.153 & 0.404$\pm$0.118 & 0.653$\pm$0.059 & 0.499$\pm$0.058 & 0.668$\pm$0.094 & 0.511$\pm$0.100 & 0.761$\pm$0.107 & 0.583$\pm$0.094 \\
        Change & /                & /                & 11.2\%          & 23.5\%          & 13.8\%          & 26.5\%          & 29.6\%          & 44.3\%          \\
        \hline
    \end{tabular}%
    }
\end{table}

\begin{figure}[htbp]
    \centering
    \includegraphics[width=0.95\linewidth]
    {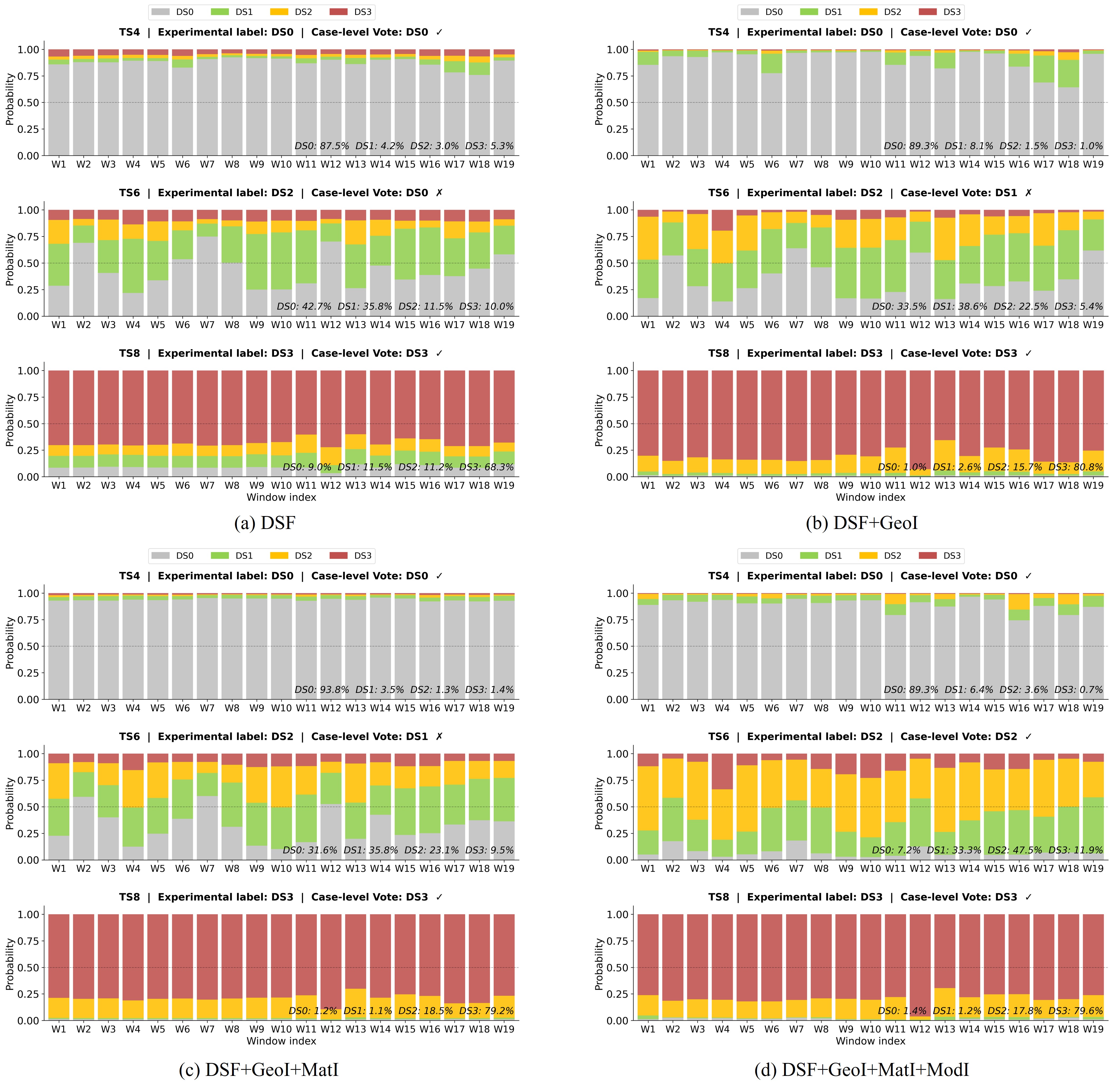}
    \caption{Window- and case-level probabilistic damage state inference for experimental WN tests TS4, TS6, and TS8. Each subplot shows the predicted DS probability vector of individual 30-s window samples under a given feature configuration. The bottom-right panel shows the mean probability vector for each test case, from which the case-level prediction is assigned by the maximum probability. Metadata enrichment progressively corrects the TS6 prediction from underestimated states toward the true DS2 label.}
    \label{case_probability}
\end{figure}

Beyond sample-level classification metrics, the experimental results were further examined at the case level to assess whether the window-based predictions led to stable DS inference for each post-earthquake WN test. Figure \ref{case_probability} presents the probabilistic outputs for TS4, TS6, and TS8. In each subplot, the bars represent the predicted probabilities of different damage states for individual 30-s window samples. The bottom-right panel summarizes the mean probability vector for each test case by averaging the window-level probabilities. The final case-level prediction was then assigned as the DS with the highest mean probability. As shown, for TS4 and TS8, the averaged probabilities are dominated by DS0 and DS3, respectively, leading to case-level predictions consistent with the capacity-curve-based labels. The main challenge appears in TS6, whose true label is DS2. With DSFs only, the case-level prediction is DS0, indicating a clear underestimation of the moderate damage condition. After adding geometrical and material information, the prediction shifts to DS1, which remains incorrect but is closer to the true state. When modal information is further included, the averaged probability becomes dominated by DS2, yielding the correct case-level prediction. This trend differs from the numerical results, where the modal information provided only a marginal additional improvement, suggesting that modal information is particularly useful in the experimental domain as it helps compensate for simulation-to-experiment discrepancies caused by differences in geometrical configurations, boundary conditions, mass distribution, stiffness characteristics, and measurement uncertainty. From a practical SHM perspective, the results indicate that post-earthquake damage assessment should not rely solely on isolated window-level classifications or DSF-only representations. Instead, aggregating probabilistic predictions over multiple response windows and incorporating readily available structural metadata and modal information can lead to more stable and physically consistent case-level DS inference. Such a strategy is particularly important for field applications, where monitoring data are often noisy, structural information is incomplete, and the target structure may deviate from the numerical population used for model training.

The preceding analyses evaluate the ability of the proposed feature representations to infer damage states across numerical and experimental structures. The following analysis addresses a separate but connected question: whether differences in the timeliness and reliability of these inferred states can materially affect recovery outcomes. The recovery model is therefore used as a consequence-propagation layer, translating classification uncertainty into inspection and repair delays under explicitly stated illustrative assumptions.

\subsection{Community resilience analysis}
Building on the damage-classification results, the previously defined SHM observation matrix was derived from the overall mean confusion matrix of the fully metadata-enriched DSF-based representation, as shown in Figure \ref{metadata_confusion}(c). This matrix was used to generate monitoring-informed damage tags for monitored buildings and propagate classification uncertainty into the community recovery simulation. The SHM-informed scenario was then compared with a baseline inspection-based scenario without SHM, while keeping the repair prioritization rule, repair resources, and damage-dependent repair durations identical. Therefore, differences in recovery outcomes can be primarily attributed to the timeliness and reliability of the available damage information, linking sample-level classification performance to community-level resilience impact.

\begin{figure}[htbp]
    \centering
    \includegraphics[width=1.0\linewidth]
    {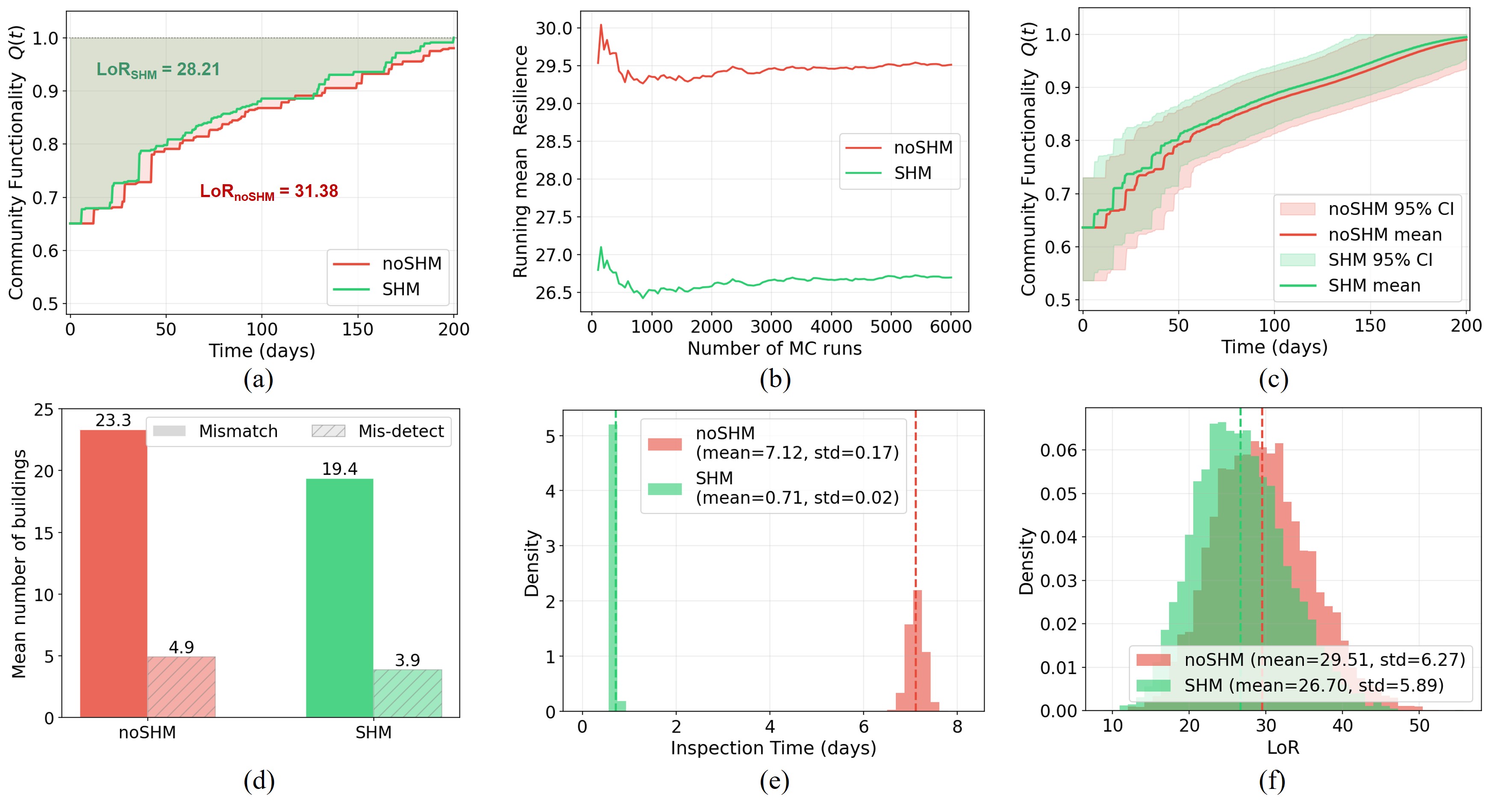}
    \caption{Community-level recovery comparison between the baseline inspection-based scenario without SHM and the SHM-informed scenario: (a) functionality recovery trajectory from one Monte Carlo realization; (b) convergence of the Monte Carlo estimate of resilience loss; (c) mean community functionality recovery curves with 95\% confidence intervals; (d) average numbers of damage state mismatches and missed detections; (e) distribution of inspection duration; (f) distribution of loss of resilience (LoR).}
    \label{community_resilience}
\end{figure}

Based on the community simulation setup described in Section 2.6, uncertainties in DS realization, damage identification, and recovery processes were explicitly considered and propagated through Monte Carlo (MC) simulations. Figure \ref{community_resilience} presents the resulting recovery responses under the earthquake scenario with PGA = 0.8 g. For the MC realization shown in Figure \ref{community_resilience}(a), the SHM-informed scenario achieves faster functionality recovery and a lower loss of resilience with \(LoR\) reduced from 31.38 to 28.21. To ensure statistical stability, repeated MC simulations were performed, and Figure \ref{community_resilience}(b) shows that the estimated mean resilience loss stabilizes after approximately 5000 runs; therefore, the subsequent statistical comparisons are based on 5000 MC realizations. The mean recovery curves with 95\% confidence intervals in Figure \ref{community_resilience}(c) further confirm that SHM-informed damage tagging leads to higher community functionality over the recovery horizon, with the improvement being more evident during the early recovery stage. This advantage is primarily associated with the acceleration of post-earthquake information acquisition. The average inspection time decreases from 7.12 days in the no-SHM scenario to 0.71 days in the SHM-informed scenario, as shown in Figure \ref{community_resilience}(e), allowing repair activities to be initiated earlier. In addition, Figure \ref{community_resilience}(d) shows that SHM-informed tagging reduces both DS mismatches and missed detections compared with conventional inspection, thereby decreasing the likelihood of repair crews being assigned based on inaccurate damage information. These improvements in information timeliness and reliability translate into a clear resilience benefit. As shown in Figure \ref{community_resilience}(f), the \(LoR\) distribution shifts toward lower values under the SHM-informed scenario, with the mean \(LoR\) decreasing from 29.51 to 26.70 by about 10\%. The slightly smaller dispersion also indicates more stable recovery outcomes across MC realizations. These results indicate that even imperfect SHM-based damage classification can provide actionable information for post-earthquake recovery scheduling, enabling earlier repair mobilization, more effective resource allocation, and thereby improving community-level resilience when classification uncertainty is explicitly propagated into the recovery decision process.

\begin{figure}[htbp]
    \centering
    \includegraphics[width=0.55\linewidth]
    {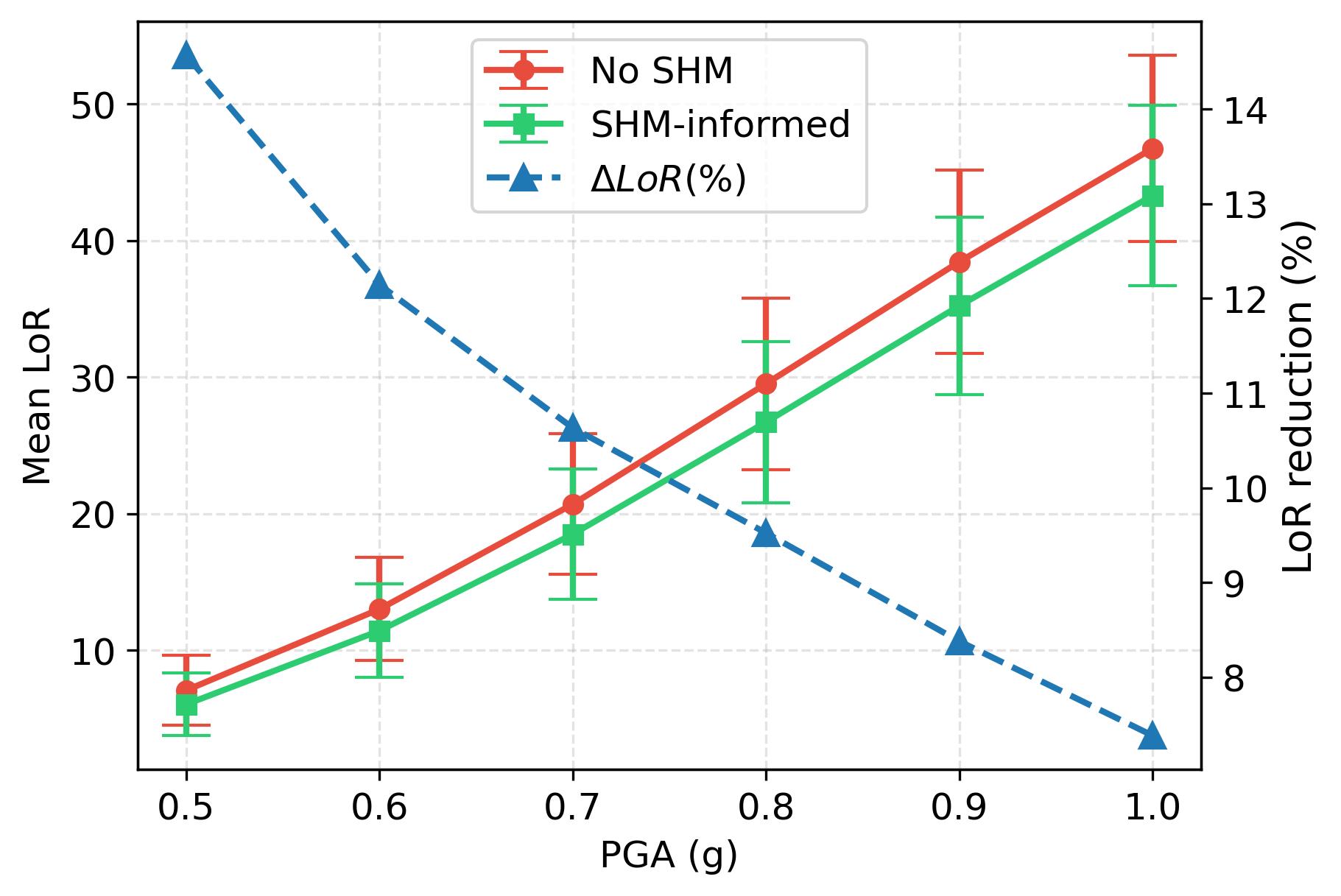}
    \caption{Influence of seismic intensity on community resilience loss. Mean LoR values with standard deviations are compared between the baseline no-SHM scenario and the SHM-informed scenario under different PGA levels. The right axis shows the relative LoR reduction achieved by SHM-informed damage tagging.}
    \label{PGA_LOR}
\end{figure}

To further examine whether the observed resilience benefit is specific to the selected PGA = 0.8 g  scenario, additional recovery simulations were conducted under different seismic intensity levels. Figure \ref{PGA_LOR} compares the mean LoR values of the no-SHM and SHM-informed scenarios and reports the corresponding LoR reduction ratios. As expected, the mean LoR increases with PGA in both scenarios since stronger shaking leads to more severe initial damage and longer recovery processes. Nevertheless, the SHM-informed scenario consistently produces lower LoR across all PGA levels, confirming the robustness of the monitoring-informed recovery benefit. The corresponding percentage reduction in LoR decreases from approximately 14.6\% at PGA=0.5 g to about 7.3\% at PGA=1.0 g. This indicates that the relative benefit of SHM-enabled damage tagging is more pronounced under moderate damage conditions, where improved damage information can more effectively influence recovery prioritization. Under more severe scenarios, widespread damage and stronger repair resource constraints limit the relative improvement achievable through damage information alone. Given that moderate-intensity earthquakes are generally more frequent than rare extreme events, these results highlight the practical value of SHM-informed damage tagging for post-earthquake assessment and recovery planning across a range of seismic scenarios.

\section{Conclusions}
This study developed a physics-informed feature fusion and structural metadata integration framework for population-based post-earthquake damage classification and resilience-oriented assessment. A heterogeneous population of nonlinear RC frame structures was generated, and structure-specific damage states were defined through pushover-based capacity interpretation rather than fixed drift thresholds. Based on sparse ground and roof acceleration measurements, physics-informed DSFs, generic time-series descriptors, convolution kernel-based representations, and metadata-enriched feature spaces were systematically evaluated under group-wise cross-structure validation. The trained classifiers were further examined using ODD shake table test data and integrated into a community-level recovery simulation to assess their downstream value for post-earthquake resilience. The key findings are summarized as follows:

First, individual physics-informed DSFs exhibited clear damage sensitivity, especially first-mode transmissibility-based features related to spectral-area variation and peak-frequency shift. However, their standalone classification capability remains limited for heterogeneous structural populations, especially for intermediate damage states. ML-based fusion of multiple DSFs substantially improved classification performance, confirming that different physically motivated features capture complementary aspects of earthquake-induced structural degradation.

Second, Physics-informed DSFs consistently outperformed Catch22 and MiniRocket under the sparse sensing and cross-structure validation setting considered in this study, suggesting that incorporating structural mechanics knowledge, such as input-output transmissibility, energy redistribution, response amplification, and stiffness-related changes, into feature design provides stronger discriminative capability than purely generic statistical or data-driven temporal representations for seismic damage identification across heterogeneous structures.

Third, structural metadata improved cross-structure generalization by providing context for interpreting response-based DSFs. Geometrical and material descriptors supplied information on structural configuration and capacity, while modal information showed a domain-dependent role. Its contribution was marginal in the numerical dataset, probably because modal-band information was already partly embedded in the transmissibility-based DSFs. In contrast, modal information markedly improved experimental validation, indicating that global dynamic properties are valuable for bridging simulation-to-experiment discrepancies caused by modeling idealization, boundary-condition uncertainty, and measurement noise.

Finally, the community-level recovery analysis showed that improved damage classification can translate into measurable resilience gains. By shortening inspection delays and reducing damage state mismatches and missed detections, SHM-informed assessment supported more effective repair prioritization and reduced resilience loss across different seismic intensity levels. These results shift the role of PBSHM from damage identification alone toward actionable decision support, where monitoring data are converted into timely information for post-earthquake recovery planning.

From a practical perspective, reliable population-based seismic damage assessment should combine physically interpretable vibration features, structural attributes, and global dynamic properties. Geometrical metadata can be obtained from design drawings, BIM models, or field surveys; material and loading information may be derived from design documents, non-destructive evaluation tests, or load assessment; and modal properties can be identified from ambient vibration measurements collected by SHM sensors. Such hybrid information provides the structural context needed to interpret vibration-derived damage indicators across heterogeneous building portfolios. Practical deployment, however, still requires careful attention to sensor layout, data quality, incomplete documentation, uncertainty in material and loading information, and the consistency between numerical training datasets and real monitored structures.

Several limitations should be acknowledged. The numerical population is restricted to two-dimensional RC frames governed primarily by flexural nonlinearities, and the feature extraction procedure relies on paired healthy and post-earthquake responses under repeatable low-amplitude excitation. The external evaluation involves one laboratory specimen and three post-earthquake structural conditions, and therefore provides an initial demonstration rather than comprehensive experimental validation. In addition, the community-recovery results depend on assumed inspection, repair, resource, and penalty parameters and should be interpreted as an illustration of consequence propagation rather than a calibrated prediction for a specific community. These limitations define the scope within which the reported transferability and resilience-related conclusions should be interpreted.

Future work should extend the proposed framework to more diverse structural typologies, three-dimensional buildings, and field monitoring datasets. Further efforts are also needed to incorporate uncertainty-aware classifiers, investigate domain adaptation strategies with limited target-domain data, and explicitly account for sensor noise, missing data, and incomplete metadata. Another important direction is cost-informed value-of-information analysis, where SHM installation, maintenance, data processing, inspection, and repair costs are jointly considered with resilience benefits. In this context, coupling PBSHM-based damage inference with more detailed recovery optimization, occupancy decision-making, and regional infrastructure interdependency models would further strengthen the role of SHM as an enabling technology for resilience-informed earthquake risk management.

\section*{Data Availability}
The code and datasets used in this study will be deposited in a public GitHub repository. The URL will be included in the published version of this article.

\section*{Acknowledgments}
The research was funded by the European Union under the program HORIZON-CL5-2023-D4-02-01 for project INBLANC - INdustrialisation of Building Lifecycle data Accumulation, Numeracy and Capitalisation, Grant Number: 101147225. Views and opinions expressed are however those of the author(s) only and do not necessarily reflect those of the European Union or CINEA. Neither the European Union nor the granting authority can be held responsible for them.

\bibliographystyle{unsrt}  
\bibliography{refs}  

\section*{Appendix}

\begin{table}[ht]
    \centering
    \caption{Details of the selected earthquake records}
    \label{tab:EQs}
    \begin{tabular}{ccccccc}
        \toprule
        No. & Waveform ID & Earthquake Event & Year & Mw & Distance (km) & PGA (m/s$^{2}$) \\
        \midrule
        1  & 55   & FRIULI  & 1976 & 6.5 & 23 & 3.49 \\
        2  & 171  & BASSOT  & 1978 & 6.0 & 18 & 1.49 \\
        3  & 199  & MONTEN  & 1979 & 6.9 & 16 & 3.68 \\
        4  & 334  & ALKION  & 1981 & 6.6 & 19 & 2.83 \\
        5  & 5651 & BANJAL  & 1981 & 5.7 & 7  & 3.55 \\
        6  & 361  & UMBRIA  & 1984 & 5.6 & 19 & 2.04 \\
        7  & 6131 & IONIAN  & 1988 & 4.8 & 12 & 2.70 \\
        8  & 948  & SICILI  & 1990 & 5.6 & 24 & 2.48 \\
        9  & 559  & PYRGOS  & 1993 & 5.4 & 25 & 1.12 \\
        10 & 6115 & KOZANI  & 1995 & 6.5 & 17 & 2.03 \\
        11 & 591  & UMBRIA  & 1997 & 5.7 & 3  & 3.38 \\
        12 & 1715 & ANOLIO  & 1999 & 6.0 & 14 & 3.20 \\
        \bottomrule
    \end{tabular}
\end{table}

\begin{table}[htbp]
\centering
\caption{Hyperparameter search spaces used for Optuna-based optimization of the investigated ML classifiers.}
\label{tab:ml_search_space}
\renewcommand{\arraystretch}{1.15}
\small
\begin{tabular}{lll}
\toprule
Classifier & Hyperparameter & Search space \\
\midrule
LR 
& \(C\) & \(10^{-3}\)--\(10^{2}\) (log) \\
& penalty & \{L1, L2\} \\
\midrule
RF 
& \(n_{\mathrm{estimators}}\) & 100--500 \\
& max depth & 3--20 \\
& min samples split & 2--20 \\
& min samples leaf & 1--10 \\
& max features & \{sqrt, log2\} \\
\midrule
XGBoost 
& \(n_{\mathrm{estimators}}\) & 100--500 \\
& max depth & 3--10 \\
& learning rate & 0.01--0.30 (log) \\
& subsample & 0.6--1.0 \\
& colsample by tree & 0.5--1.0 \\
& min child weight & 1--15 \\
& reg alpha & \(10^{-8}\)--10 (log) \\
& reg lambda & \(10^{-8}\)--10 (log) \\
\midrule
SVM 
& kernel & \{RBF, polynomial, sigmoid\} \\
& \(C\) & 0.01--100 (log) \\
& \(\gamma\) & \(10^{-4}\)--10 (log) \\
& degree & 2--5, for polynomial kernel \\
\midrule
MLP 
& hidden layers & 1--3 \\
& units per layer & 16--256 \\
& activation & \{ReLU, tanh\} \\
& learning rate init & \(10^{-4}\)--0.1 (log) \\
& \(\alpha\) & \(10^{-5}\)--0.1 (log) \\
\midrule
KNN 
& \(n_{\mathrm{neighbors}}\) & 3--30 \\
& weights & \{uniform, distance\} \\
& \(p\) & \{1, 2\} \\
\bottomrule
\end{tabular}
\end{table}

\end{document}